\documentclass[%
 reprint,
nofootinbib,
 amsmath,amssymb,
 aps,
prd,
]{revtex4-1}

\usepackage{amssymb,amsmath}
\usepackage{euscript}
\usepackage{mathtools} 
\usepackage[usenames,dvipsnames]{color}
\usepackage{graphicx}
\usepackage{dcolumn}
\usepackage{bm}
\usepackage{hyperref}
\usepackage{pifont} 
\usepackage{mathrsfs}  

\newcommand{\C}{\EuScript{C}}

\newcommand{\betahat}{\hat{\bm \beta}}

\newcommand{\gammahat}{\hat{\bm n}}
\newcommand{\Khat}{\bm{\widehat{K}}}

\newcommand{\cmark}{\ding{51}}
\newcommand{\xmark}{\ding{55}}

\begin{document}

\preprint{APS/123-QED}

\title{A Generalized Doppler and Aberration Kernel\\ for Frequency-Dependent Cosmological Observables}

\author{Siavash Yasini and Elena Pierpaoli}
\affiliation{%
 Physics \& Astronomy Department, University of Southern California, Los Angeles, California,  90089-0484 \\
}%

\date{\today}

\begin{abstract}
We introduce a \emph{frequency-dependent} Doppler and aberration transformation kernel for the harmonic multipoles of a general cosmological observable with spin weight $s$, Doppler weight $d$ and arbitrary frequency spectrum. In the context of Cosmic Microwave Background (CMB) studies, the frequency-dependent formalism allows to correct for the motion-induced aberration and Doppler effects on individual frequency maps with different masks. It also permits to deboost background radiations with non-blackbody frequency spectra, like extragalactic foregrounds and CMB spectra with primordial spectral distortions. The formalism can  also be  used to correct  individual E and B polarization modes and account for motion-induced E/B mixing of polarized observables with $d\neq1$ at different frequencies. 

We apply the generalized aberration kernel on polarized and unpolarized specific intensity at 100 and 217 GHz and show that the motion-induced effects typically increase with the frequency of observation. In all-sky CMB experiments, the frequency-dependence of the motion-induced effects for a blackbody spectrum are overall negligible. However in a cut-sky analysis, ignoring the frequency dependence can lead to percent level error in the polarized and unpolarized power spectra over all angular scales. 
In the specific cut-sky used in our analysis ($b > 45^\circ, f_\text{sky}\simeq14\%$), and for the dipole-inferred velocity $\beta=0.00123$ typically attributed to our peculiar motion, the Doppler and aberration effects can change polarized and unpolarized power spectra of specific intensity in the CMB rest frame by $1-2\%$, but we find the polarization cross-leakage between E and  B modes to be negligible. 
\end{abstract}

\maketitle


\section{Introduction}
Our local peculiar motion with respect to the cosmic microwave background (CMB) imprints an effect on the observed frequency and angle of the incoming photons. Consequently, the motion affects the observed spherical harmonic multipoles ($\ell$ modes) of the CMB, which are in turn used for cosmological parameter estimation and other statistical analyses. For a statistically isotropic sky, the harmonic multipoles of the CMB are uncorrelated with each other in the CMB rest frame. However, this is not the case in a frame that is moving with respect to the CMB. In a moving frame the nearby multipoles leak into each other and hence couple together (i.e. they become correlated) \cite{Challinor2002}. This correlation, if not accounted for, can generate a bias on the estimated parameters inferred from observations \cite{Catena:2012hq}.

Correction of the local boost effects on the measured cosmological data (\emph{deboosting}) can indeed be performed in real space. However the real space approach presents some numerical challenges. It has been shown that  deboosting the CMB in real space can lead to a spurious power suppression over small scales ($\ell > 1000$) and in order to correct for it, very high resolution is needed \cite{Yoho:2012am}. This in turn calls for an increase in the number of pixels involved and makes the process computationally expensive. The harmonic space deboosting on the other hand, does not manifest this problem and is more numerically efficient \cite{Jeong:2013sxy}. Moreover, pixel window functions, beam and mask effects can be more easily dealt with in harmonic space and it also allows us to apply the motion-induced effects directly on the power spectrum. Therefore, studying the motion-induced effects on the CMB in harmonic space is of practical advantage. 


The transformation of the spherical harmonic multipoles  of the CMB from one moving frame to another is typically referred to as the \emph{Doppler and aberration kernel} (or more commonly just \emph{aberration kernel}). The first calculation of the aberration kernel for integrated intensity in an all-sky experiment was first presented in Ref.  \cite{Challinor2002}, which employs a Taylor expansion in the dimensionless frame velocity $\beta\equiv v/c$. Since the effect of Doppler and aberration on harmonic modes grows roughly as $\propto \beta \ell$, this perturbative scheme is only valid for harmonic multipoles up to $\ell \simeq \beta^{-1}$. For the value of the dipole-inferred velocity with respect to the CMB $\beta=0.00123$ \cite{Kogut:1993ag}, this corresponds to $\ell \lesssim 800$. The  authors report the motion-induced effects on the temperature power spectrum to be of order $\beta^2\simeq 10^{-6}$ for an all-sky map and hence negligible.

The study of the Doppler and aberration effects were then extended to higher $\ell$ modes using real space boosting \cite{Yoho:2012am} and some innovative techniques for calculating the aberration kernel such as Legendre polynomial recursive relations \cite{Chluba2011} and Bessel function fits \cite{Notari:2011sb}. 
Later, Ref. \cite{Dai2014} (DC from here on) introduced new recursive relations that allow to calculate the kernel for observables with different Doppler weights (see \S\ref{sec:II}), along with an ingenious system of ordinary differential equations (ODE) that remarkably reduces the calculation time of the kernel elements at very high $\ell$. The all- and masked-sky analyses of the boost on the CMB reveal that the motion-induced effects generally become larger at smaller angular scales, and can reach  $\gtrsim 1\%$ in power spectrum amplitude for certain sky cuts \cite{Jeong:2013sxy}.

The correlation between nearby CMB multipoles due to the motion-induced leakage is also important for dipole-independent measurements of our local velocity with respect to the CMB frame \cite{Kosowsky2010,Amendola2010,Burles:2006xf,Notari:2011sb,Yasini:2016dnd,Notari:2015daa}. Using a wide range of $\ell$ modes, the \emph{Planck} team has measured this dipole-independent velocity to be $\beta=0.00128\pm0.00026(stat.)\pm0.00038(syst.)$ \cite{Aghanim:2013suk}, which is consistent with the dipole-inferred value, but with relatively large error bars. The next generation of the CMB surveys are certainly going to make a more precise measurement of this motion-induced correlation, but in order to do so, it is essential to have a precise calculation of the Doppler and aberration kernel elements for the harmonic modes. Therefore, not only the aberration kernel calculations are essential for boost \emph{correction}, they can be used for boost \emph{detection} as well \cite{Chluba2011}.  

So far, the development of the aberration kernel has been mostly aimed at frequency-independent observables, such as thermodynamic temperature or the integrated (bolometric) intensity of an initial blackbody spectrum \footnote{Exceptions to this include the frequency-dependent calculations presented in the appendices of Refs. \cite{Challinor2002,Chluba:2012gq,Yasini:2016dnd}. However, contrary to the methods introduced in this paper, those calculations are only valid for $\beta \ell \ll 1$.}. Therefore, the available formulas cannot be readily applied to frequency-dependent observables (e.g. specific intensity $I_\nu$). A frequency-dependent aberration kernel allows for correction of the boost in the observed harmonic multipoles of an individual frequency map of any background radiation with an arbitrary frequency spectrum. This is of practical importance because it allows  to apply different masks at each observational frequency, and appropriately correct for the motion-induced effects in the power spectra of individual maps before combining them for final parameter estimation. Additionally, using a frequency-dependent aberration kernel we can cross-correlate different frequency maps with their respective masks, and enhance the detection significance of the dipole-independent boost measurements. 

It is important to have an unambiguous measurement of the motion-induced correlation between the harmonic multipoles of the CMB and separate them from other sources of mode coupling (e.g. primordial non-Gaussianity \cite{Catena:2013qd}) in both polarized and unpolarized components. Correcting the boost effects in the CMB polarization is arguably even more important than the unpolarized radiation, due to their relevance to the detection of the primordial gravitational waves. Since the current and future generation of  CMB polarization surveys perform in different frequency bands, it is crucial to have a frequency-dependent formalism for correcting the boost effects in the measurements of these CMB experiments. 

In this paper, we introduce a generalized Doppler and aberration kernel operator which can be applied to a general frequency-dependent observable with arbitrary Doppler and spin weight (\S \ref{sec:II}). Our calculations heavily rely on the formulas derived in DC \cite{Dai2014}. We apply the generalized Doppler and aberration kernel to both unpolarized and polarized pure blackbody CMB maps and power spectra (\S \ref{sec:III}), and show that the distinction between the generalized and DC kernel elements is not observationally significant for all-sky maps, but it cannot be neglected for masked skies. Even though the focus of the paper is the application of the aberration kernel in CMB observations, it can also be applied to any frequency-dependent cosmological observable (e.g. CIR, CRB etc.) or in radiative transfer scattering problems.

The paper is organized as follows: In \S \ref{sec:II} we layout the theoretical development of the generalized Doppler and aberration kernel and introduce a frequency-dependent formula for boosting/deboosting harmonic multipoles of cosmological observables.  In \S \ref{sec:IIIA}(1-2) we analyze the behavior of the aberration kernel elements for unpolarized CMB temperature, integrated intensity and specific intensity at different frequencies. In \S \ref{sec:IIIA}(3-4) we repeat the analysis for polarized CMB radiation and compare the results to the unpolarized case. In \S \ref{sec:IIIB} we apply the aberration kernel to both polarized and unpolarized CMB power spectra and study the frequency dependence of the motion-induced effects in an all-sky experiment and its observational implications.  We repeat our analysis in \S \ref{sec:IIIC} for a masked-sky and show that the effects become more relevant in this case. Finally we summarize the results in In \S \ref{sec:IV}. 

\section{Generalized Doppler and aberration kernel}\label{sec:II}
In this section we present the calculations of the Doppler and aberration kernel for a frequency-dependent observable. We consider a general observable $_s^dX(\nu,\gammahat)$ in the direction $\gammahat$ at the frequency $\nu$ with a spin weight $s$ and Doppler weight $d$. The spin weight determines how the observable transforms under coordinate rotation. Scalar, vector and tensor observables respectively have spin weights of $0, \pm1$ and  $\pm2$. On the other hand, the Doppler weight determines how the observable transforms under a Lorentz boost. For an observable with Doppler weight $d$, the quantity $_s^dX/\nu^d$ is a Lorentz invariant.  Note that the Doppler weight of the observable does not necessarily correlate with its frequency dependence. As a simple example, for a pure blackbody radiation the parameters $I_\nu$ (specific intensity) and $T^3$ (thermodynamic temperature cubed) both have Doppler weights of 3, with the former being frequency-dependent but not the latter. Table \ref{tab:1} lists a few common observable and their respective Doppler and spin weights. 
\begin{table*}
	\centering
	\begin{tabular}{|c|c|c|c|c|} 
		\hline
		Observable &  Notation &  Doppler Weight & Spin Weight & Frequency Dependent \\
		\hline
		\hline
		Thermodynamic Temperature &   $T$  & 1 & 0 & \xmark\\
		\hline
		
		Specific Intensity & $I_\nu$  & 3 & 0 & \cmark\\
		\hline
		Integrated (bolometric) Intensity & $I = \int I_\nu \rm d \nu$ & 4 & 0 & \xmark \\
		\hline
		
		Polarized Temperature & $Q_T\pm i U_T$ & 1 & $\mp2$ & \xmark \\
		\hline
		Polarized Intensity & $Q_\nu \pm i U_\nu$ & 3 & $\mp 2$ & \cmark  \\
		\hline
		Unpolarized Occupation Number & $n_\nu$ & 0 & 0 & \cmark \\
		\hline
	\end{tabular}
	\caption{Some common observables used in CMB analysis. The polarization  parameters $Q$ and $U$ can conventionally have opposite spin signs. The more common polarization parameters $E$ and $B$ are not listed, because they have mixed spin weights.  }
	\label{tab:1}
\end{table*}

In a frame moving with the velocity $\beta=v/c$ in the $\betahat$ direction, $_s^dX(\nu,\gammahat)$ will be observed at a different frequency $\nu'$ and direction $\gammahat'$ due to the Doppler and aberration effects. Using the Lorentz invariance of $_s^dX/\nu^d$ we can write \cite{Rybicki1986,Ford:2013koa}

\begin{equation}\label{X_lorentz_transform}
_s^d\tilde{X}({\nu'},\gammahat') = \left( \frac{\nu'}{\nu} \right)^d ~_s^dX(\nu,\gammahat),
\end{equation}
 with 
 \begin{equation}\label{doppler_freq}
 \nu'=\gamma (1+\beta \mu)\nu,
 \end{equation}
 and
 \begin{equation}\label{aberration}
 \gammahat'=\Big(\frac{(1-\gamma^{-1})\mu+\beta}{1+\beta \mu}\Big )\betahat+\Big (\frac{\gamma^{-1}}{1+\beta \mu}\Big)\gammahat,
 \end{equation}
 where  $\gamma=1/\sqrt{1-\beta^2}$ and $\mu=\gammahat \cdot \betahat$.  Equations \eqref{doppler_freq}  and \eqref{aberration} respectively represent the Doppler  and aberration effects. For simplicity, we assume that $\betahat=\hat{\bm z}$ and rewrite equation \eqref{aberration} as 
 \begin{equation}\label{aberration_z}
 \mu'=\frac{\mu+\beta}{1+\beta \mu}
 \end{equation}
 where $\mu'=\gammahat' \cdot \betahat$. It is worth mentioning that in practice, we will be using the inverse of the Doppler and aberration relations (Eq. \eqref{doppler_freq} and \eqref{aberration_z})
 
 \begin{equation}\label{doppler_freq_inv}
 \nu=\gamma (1-\beta \mu')\nu',
 \end{equation}
 
 \begin{equation}\label{aberration_z_inv}
 \mu=\frac{\mu'-\beta}{1-\beta \mu'}.
 \end{equation}
  
 We are interested in finding how the harmonic multipoles of $_s^dX(\nu,\gammahat)$ transform under a boost, so we expand both sides of equation \eqref{X_lorentz_transform} using 
 
 \begin{equation}\label{X_expansion}
 _s^dX(\nu,\gammahat) = \sum_{\ell=0}^{\infty}\sum\nolimits_{m}^{\ell}~_s^dX_{\ell m}(\nu)~ _{-s}Y_{\ell m}(\gammahat).
 \end{equation}

By separating the coefficients on the left hand side of Eq. \eqref{X_lorentz_transform} we obtain

\begin{multline}\label{X'lm_kernel}
_s^d\tilde{X}_{\ell' m'}(\nu') =\\
 \sum_{\ell,m} \int \text{d} \gammahat'^2 ~ _{-s}Y^*_{\ell' m'}(\gammahat')~_{-s}Y_{\ell m}(\gammahat) \left( \frac{\nu'}{\nu} \right)^d ~_s^dX_{\ell m}(\nu) =\\
 \sum_{\ell,m} \int \text{d} \gammahat'^2   \frac{_{-s}Y^*_{\ell' m'}(\gammahat')_{-s}Y_{\ell m}(\gammahat)}{[\gamma (1-\beta \mu')]^d} ~_s^dX_{\ell m}(\nu).
\end{multline}
This expression is usually represented as a \emph{harmonic boost equation} \cite{Challinor2002}

\begin{equation}\label{X'lm_kernel_short}
_s^d\tilde{X}_{\ell' m'}(\nu') =
\sum_{\ell,m}~_s^d\mathcal{K}^{m' m}_{\ell' \ell}(\beta) ~_s^dX_{\ell m}(\nu),
\end{equation}
where the aberration kernel $~_s^d\mathcal{K}^{m' m}_{\ell' \ell}(\beta)$ represents the angular integral in equation \eqref{X'lm_kernel}

\begin{equation}\label{Klm_definition}
_s^d\mathcal{K}^{m' m}_{\ell' \ell}(\beta)=
 \int \text{d} \gammahat'^2   \frac{_{-s}Y^*_{\ell' m'}(\gammahat')_{-s}Y_{\ell m}(\gammahat)}{[\gamma (1-\beta \mu')]^d}.
\end{equation}

 For $\betahat=\hat{\bm z}$ different $m$ modes do not mix, so we will drop the index $m'$ and use the notation $~_s^d\mathcal{K}^{m}_{\ell' \ell}(\beta)=\sum_{m'}  \delta_{m' m}~_s^d\mathcal{K}^{m' m}_{\ell' \ell}(\beta)$. From here on, we will refer to $~_s^d\mathcal{K}^{m}_{\ell' \ell}(\beta)$ as DC kernel elements. 

The abbreviation of Eq. \eqref{X'lm_kernel} as Eq. \eqref{X'lm_kernel_short} is only correct if the observable $_s^dX_{\ell m}$ is frequency-independent and it is therefore only applicable to observables marked with a cross in Table \ref{tab:1}. Eq. \eqref{X'lm_kernel_short} implicitly assumes that the frequency dependence of the observable in the moving frame, $_s^d\tilde{X}_{\ell'm}(\nu')$, is the same as the one in the rest frame, $_s^dX_{\ell m}(\nu)$. For a general frequency-dependent observable, since the argument of $_s^dX_{\ell m}(\nu)$ implicitly depends on $\gammahat'$ (see Eq. \eqref{doppler_freq_inv}), it cannot be taken out of the integral in Eq. \eqref{X'lm_kernel}. In other words, the frequencies $\nu'$ observed in the moving frame are angle dependent, and are not the same as the frequencies observed in the rest frame. In order to calculate the aberration kernel for a frequency-dependent observable, we will have to integrate over this implicit angular dependence. As we will show, this procedure (generalization) will result in frequency-dependent kernel elements.

In order to implement the frequency dependence of Eq. \eqref{X'lm_kernel}, we need to expand $~_s^dX_{\ell m}(\nu) $ about the observed frequency in the moving frame $\nu'$. Using Taylor expansion we can write
\begin{multline}\label{X_lm-frequency_Taylor}
~_s^dX_{\ell m}(\nu) = \sum_{n=0}^{\infty}\frac{(\nu-\nu')^n}{n!} \partial^n_{\nu'} ~_s^dX_{\ell m}(\omega)\\
= \sum_{n=0}^{\infty}\frac{(\gamma(1-\beta \mu')-1)^n}{n!} \nu'^n\partial^n_{\nu'} ~_s^dX_{\ell m}(\omega),\\
\end{multline}
where $\partial^n_{\nu'}$ is shorthand notation for $\partial^n/\partial_{\omega}\big| _{\omega = \nu'}$ and $\omega$ is a dummy variable representing the frequency. Here, in the second line we used $\nu = \gamma(1-\beta \mu')\nu'$. Now using the binomial expansion 
\begin{equation}
(\gamma(1-\beta \mu')-1)^n = \sum_{k=0}^{n} \binom{n}{k}(-1)^{n+k}[\gamma(1-\beta \mu')]^k,
\end{equation}
we can rewrite equation \eqref{X_lm-frequency_Taylor} as 

\begin{multline}\label{X_lm-frequency_Taylor2}
~_s^dX_{\ell m}(\nu) = \\
 \sum_{n,k}\frac{(-1)^{n+k}}{n!} \binom{n}{k}[\gamma(1-\beta \mu')]^k\nu'^n\partial^n_{\nu'} ~_s^dX_{\ell m}(\omega).
\end{multline}
Substituting this into Eq. \eqref{X'lm_kernel} yields 

\begin{multline}\label{X'lm_kernel_short2}
_s^d\tilde{X}_{\ell' m'}(\nu') =\\
\sum_{\ell,m} \sum_{n,k} \frac{(-1)^{n+k}}{n!} \binom{n}{k}~_{~~~s}^{d-k}\mathcal{K}^{m}_{\ell' \ell}(\beta) \nu'^n\partial^n_{\nu'} ~_s^dX_{\ell m}(\omega).
\end{multline}
Depending on the expansion order, the coefficients $~^{d-k}_{~~~s}\mathcal{K}^{m}_{\ell' \ell}(\beta)$ can be calculated using the recursive formulas presented in DC \cite{Dai2014}

\begin{multline}\label{recursive_1}
~^{d}_{s}\mathcal{K}^{m}_{\ell' \ell}=
 \gamma ~^{d-1}_{~~~s}\mathcal{K}^{m}_{\ell' \ell} + \gamma \beta \Big[ ~_s\C^m_{\ell+1}  ~^{d-1}_{~~~s}\mathcal{K}^{m}_{\ell' \ell+1} \\
 + \frac{s m }{\ell (\ell+1) } ~^{d-1}_{~~~s}\mathcal{K}^{m}_{\ell' \ell} + ~_s\C^m_{\ell}  ~^{d-1}_{~~~s}\mathcal{K}^{m}_{\ell' \ell-1} \Big],
\end{multline}
or 
\begin{multline}\label{recursive_2}
~^{d}_{s}\mathcal{K}^{m}_{\ell' \ell}=
\gamma ~^{d+1}_{~~~s}\mathcal{K}^{m}_{\ell' \ell} - \gamma \beta \Big[ ~_s\C^m_{\ell'+1}  ~^{d+1}_{~~~s}\mathcal{K}^{m}_{\ell'+1 \ell} \\
+ \frac{s m }{\ell' (\ell'+1) } ~^{d+1}_{~~~s}\mathcal{K}^{m}_{\ell' \ell} + ~_s\C^m_{\ell'}  ~^{d+1}_{~~~s}\mathcal{K}^{m}_{\ell'-1 \ell} \Big],
\end{multline}
where 

\begin{equation}\label{Clm}
_s\C^m_\ell = \begin{cases}
\sqrt{\frac{(\ell^2-m^2)(\ell^2-s^2)}{\ell^2(4\ell^2-1)}}  & \ell>0~ \&~\ell > |m|,|s|\\
0 & \text{otherwise.}\\
\end{cases}
\end{equation} 
Using Eq. \eqref{recursive_1} and \eqref{recursive_2} one can find the kernel coefficients in Eq. \eqref{X'lm_kernel_short2} in terms of $~^{1}_{s}\mathcal{K}^{m}_{\ell' \ell}$, which can  be calculated using the following system of coupled ordinary differential equations (ODE) \cite{Dai2014}:

\begin{equation}\label{ODE}
\partial_\eta ~^{1}_{s}\mathcal{K}^{m}_{\ell' \ell} = (\ell+1)~_s\C^m_{\ell+1}  ~^{1}_{s}\mathcal{K}^{m}_{\ell' \ell+1} \\
-\ell ~_s\C^m_{\ell}  ~^{1}_{s}\mathcal{K}^{m}_{\ell' \ell-1},
\end{equation} 
where $\eta=\tanh^{-1}\beta$ is the rapidity of the moving frame. This ODE system can be solved using the initial condition $~^{1}_{s}\mathcal{K}^{m}_{\ell' \ell}=\delta_{\ell' \ell}$ for $\eta=0$.

Now, we define the generalized Doppler and aberration kernel as an operator
\begin{equation}\label{Generalized_aberration_kernel}
~_{s}^{d}\Khat^{m}_{\ell' \ell}(\beta,\nu')  \equiv \sum_{n=0}^{\infty} \sum_{k=0}^{n} \frac{(-1)^{n+k}}{n!} \binom{n}{k} ~_{~~~s}^{d-k}{\mathcal{K}}^{m}_{\ell' \ell}(\beta) \nu'^n\partial^n_{\nu'}
\end{equation}
and rewrite Eq. \eqref{X'lm_kernel}
\begin{equation}
_s^d\tilde{X}_{\ell' m}(\nu') =
\sum_{\ell} ~_{s}^{d}\Khat^{m}_{\ell' \ell}(\beta,\nu')  ~_s^dX_{\ell m}(\nu).
\end{equation}
We have replaced the dummy variable $\omega$ with $\nu$ to assimilate this equation with Eq. \eqref{X'lm_kernel_short}. Note that the right hand side is only a function of $\nu'$ and $\nu$ will be replaced by $\nu'$ after differentiation. Now, using this generalized kernel we will study the motion-induced effects in the CMB maps. 

\section{Results}\label{sec:III}
\subsubsection{Notation}
Here we layout some of the notation used in the following section for easy reference. 
The blackbody and differential blackbody frequency functions are respectively defined as 

\begin{align}
B_\nu(T_0) &= \frac{2 h }{c^2} \frac{\nu^3}{e^{h \nu/k T_0}-1}, \\
F_\nu(T_0) &= \frac{\partial B_\nu(T)}{\partial T}\Big|_{T=T_0}=\frac{B_\nu(T_0)}{T_0}\frac{(\nu/kT_0) e^{h \nu/k T_0}}{e^{h \nu/k T_0}-1}.
\end{align}
where the $T_0=2.725$K is the mean temperature of the CMB and $h$, $k$ and $c$ are respectively the Planck constant, Boltzmann constant and the speed of light. 

The harmonic multipoles for temperature, integrated intensity and specific intensity in the rest frame are respectively represented as $a^T_{\ell m}$, $a^I_{\ell m}$ and $a^{I_{\nu}}_{\ell m}(\nu)$, where $a^X_{\ell m} =  \int d\gammahat X(\gammahat) Y^*_{\ell m}(\gammahat)$ and their counterparts in the moving frame are denoted as $\tilde{a}^T_{\ell' m}$, $\tilde{a}^I_{\ell' m}$ and $\tilde{a}^{I_{\nu}}_{\ell' m}(\nu')$, where $\tilde{a}^X_{\ell' m} =  \int d\gammahat' X(\gammahat') Y^*_{\ell' m}(\gammahat')$. Here we have used $m'=m$, since we assumed $\betahat\ = \hat{\bm z}$. Similarly, we use the notation $e^T_{\ell m}$, $e^I_{\ell m}$ and $e^{I_{\nu}}_{\ell m}(\nu)$ for the E mode multipoles of CMB polarization $P_{\pm}(\gammahat) = Q(\gammahat) \pm iU(\gammahat)$ and $b^T_{\ell m}$, $b^I_{\ell m}$ and $b^{I_{\nu}}_{\ell m}(\nu)$ for the B modes \cite{Challinor2002,Dai2014,Notari:2011sb}. 

As in the previous section, the frequency-independent DC elements for an observable with the Doppler weight $d$ and spin weight $s$ are represented with $~_{s}^{d}\mathcal{K}^{m}_{\ell' \ell}(\beta)$. The generalized aberration kernel operator for a frequency-dependent observable (as defined in Eq. \eqref{Generalized_aberration_kernel}) is represented with $~_{s}^{d}\Khat^{m}_{\ell' \ell}(\beta,\nu') $.  The frequency function of the hamonic multipoles of the specific intensity for a pure blackbody CMB at the mean temperature of $T_0$ is given by $F_\nu(T_0)$.  For easy comparison of the boosted specific intensity multipoles with  those of the thermodynamic temperature, we normalize the generalized aberration kernel using the following definition
\begin{equation}\label{normalized_kernel}
~_{s}^{d}K^{m}_{\ell' \ell}(\beta,\nu') \equiv F_{\nu'}(T_0)^{-1} ~_{s}^{d}\Khat^{m}_{\ell' \ell}(\beta,\nu')F_\nu(T_0).
\end{equation}
We will refer to this as the normalized aberration kernel (see \S \ref{sec:IIIA2}) and it can be thought of as the eigenfunctions of the generalized aberration kernel operator $~_{s}^{d}\Khat^{m}_{\ell' \ell}(\beta,\nu') $. It is very straightforward to implement the generalized aberration kernel formalism on non-blackbody spectra: one only needs to replace $F_\nu$ with the frequency spectrum of the new background radiation.

For clarity, we have employed the $\mathtt{ \\mathcal}$ font for the frequency-independent DC kernel elements $_s^d\mathcal{K}^m_{\ell' \ell}(\beta)$ and their associate boost power transfer matrix $_s^d\mathcal{B}_{\ell'\ell}(\beta)$ (see \S \ref{sec:IIIB}), and the standard font has been used for the frequency-dependent Kernel operator $~_{s}^{d}\Khat^{m}_{\ell' \ell}(\beta,\nu')$, the normalized kernel elements $~_{s}^{d}K^{m}_{\ell' \ell}(\beta,\nu')$ and their power transfer matrix $_s^dB_{\ell'\ell}(\beta,\nu')$. The frequency-dependent variables can also be distinguished by the appearance of the explicit frequency parameter $(\nu')$ in front of them. \\
We use the notation $\Delta\mathcal{K}_X$ to show the relative percent difference between whatever kernel is under discussion and the equivalent DC kernel element with $d=1$. This notation is explicitly defined in the text wherever used. To be consistent with the calculations of \S \ref{sec:II}, the primed notation $\nu'$ and $\ell'$ has been reserved for the observed frequency and angular mode in the moving frame. We also use the dipole-inferred value of $\beta=0.00123$ in the numerical calculations.  
\subsubsection{Numerical Methods}
In \S \ref{sec:IIIA} we examine the characteristics of the generalized kernel elements $~_{s}^{d}K^{m}_{\ell' \ell}(\beta,\nu')$ and the DC kernel elements $~_{s}^{d}\mathcal{K}^{m}_{\ell' \ell}(\beta)$ for temperature $(d=1)$ and integrated intensity $(d=4)$ and compare them with each other. 

In order to calculate  $~_{s}^{d}K^{m}_{\ell' \ell}(\beta,\nu')$ we take the following steps: 
\begin{enumerate}
	\item Find the DC temperature kernel elements $~_{s}^{1}\mathcal{K}^{m}_{\ell' \ell}(\beta,\nu')$ by solving the ODE system Eq. \eqref{ODE} up to $\ell_\text{max}$.
	\item Find the DC kernel elements for Doppler weights $\{d, d-1, ... d-n\}$ using Eqs. \eqref{recursive_1} and \eqref{recursive_2}, starting from $~_{s}^{1}\mathcal{K}^{m}_{\ell' \ell}(\beta,\nu')$ obtained in step 1. 
	\item Construct the generalized Doppler and aberration kernel operator  using Eq. \eqref{Generalized_aberration_kernel} and the DC kernel elements obtained in step 2.
	\item Apply the generalized aberration kernel operator to the differential blackbody spectrum according to Eq. \eqref{normalized_kernel} to find the normalized aberration kernel elements. 
\end{enumerate}
In the numerical calculations we set the velocity parameter to $\beta=0.00123$ and the mean temperature of the CMB to $T_0=2.725$ K and $\ell_\text{max}=3000$. We also choose the $\beta$-expansion order $n=5$ to attain convergence in Eq. \eqref{Generalized_aberration_kernel}. 

For the range of angular scales that we examine, $0 \leq \ell \leq3000$, the contribution of modes $|\ell' - \ell| > 8$ is negligible so we only solve the ODE for a neighborhood of $\Delta \ell =8$ to minimize the computation time. Step 1 is the most computationally expensive step in the numerical evaluation of the generalized Kernel ($\sim$10 minutes on a 3.1 GHz Intel Core i5). We only evaluate the $~_{0}^{1}\mathcal{K}^{m}_{\ell' \ell}(\beta)$ elements once, and tabulate them for repeated use in steps 2 ($\sim$ 1 minute) and 3 ($\sim$ a few seconds). 

In \S \ref{sec:IIIB} and \S \ref{sec:IIIC}  we apply the generalized and DC aberration kernel to both polarized and unpolarized simulated skies and CMB power spectra. The power spectra are generated with CAMB\footnote{\href{http://camb.info}{http://camb.info}}  \cite{Lewis:1999bs}, using the parameters $H_0=67.74$ , $\Omega_bh^2=0.022$, $\Omega_c h^2=0.1188$, $\Omega_k=0$, $\tau=0.066$, $n_s=0.9667$ and tensor to scalar ratio $r=0$ unless otherwise noted. All the boosts on the power spectra and the sky realizations generated from them are performed in harmonic space. 

\subsection{General Characteristics of the Kernel Elements}\label{sec:IIIA}

\begin{figure}[t]
	\centering
	\includegraphics[width=1.0\linewidth]{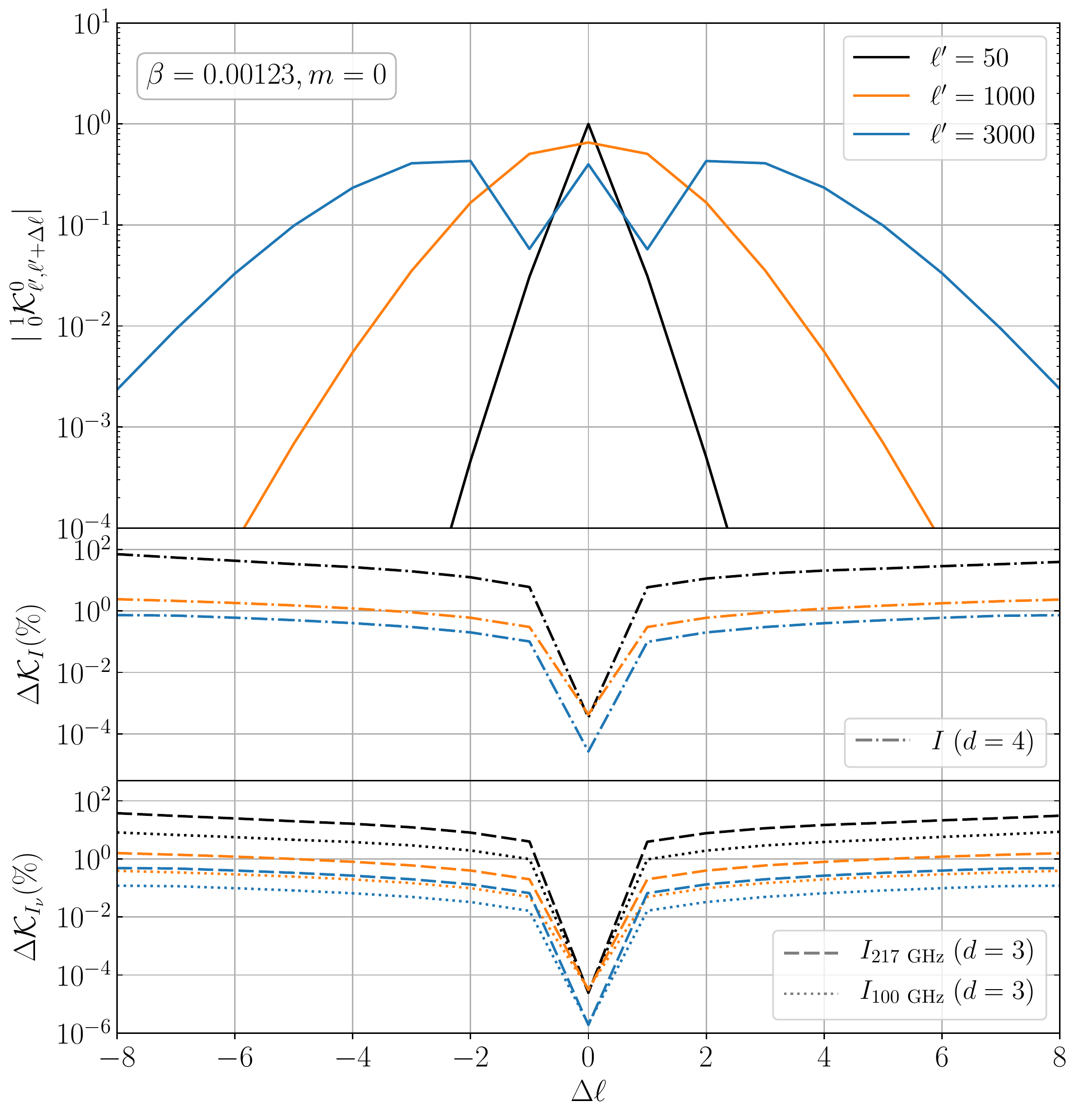}
	\caption{\emph{top}: Modulus of the temperature Doppler and aberration kernel elements for nearby multipoles of different $\ell'$ modes. Further neighbors contribute more at larger $\ell'$s. \emph{middle}: Absolute relative difference between the integrated intensity kernel (Doppler weight 4) and the temperature kernel elements of the top panel (Doppler weight 1): $\Delta \mathcal{K}_I=| ~_0^4\mathcal{K}^{0}_{\ell' \ell}(\beta)/~_0^1\mathcal{K}^{0}_{\ell' \ell}(\beta)|-1$. In general, the relative difference is larger for farther neighbors of smaller $\ell'$, but it becomes negligible as $\ell'$ grows.  \emph{bottom}: Absolute relative difference between the specific intensity kernel (Doppler weight 3) and the top panel: $\Delta \mathcal{K}_{I_{\nu}}=| ~_0^3K^{0}_{\ell' \ell}(\beta,\nu')/~_0^1\mathcal{K}^{0}_{\ell' \ell}(\beta)|-1$ for two different frequencies. The difference is is typically larger at higher frequencies and for lower $\ell$ values. Also see Fig. \ref{fig:kernel_nu_delta1}.}
	\label{fig:kernel_m_0}
\end{figure}

\subsubsection{Thermodynamic Temperature and Integrated Intensity}\label{sec:IIIA1}
Before we examine the frequency dependence of the motion-induced effects on specific intensity, first we apply the generalized aberration kernel to the well studied cases of thermodynamic temperature and integrated intensity \cite{Challinor2002,Chluba2011} and analyze the general behavior of the kernel over different angular scales and its dependence on the Doppler weight of the observable. The common way to analyze the CMB data is to convert the observed intensity $I_\nu$ in every direction to a thermodynamic temperature $T$, using the blackbody spectrum $B_\nu(T) = \frac{2 h }{c^2} \frac{\nu^3}{e^{h \nu/k T}-1}$. Obviously, the underlying assumption here is that the frequency spectrum of the CMB is Planckian in every direction: $I_\nu(\gammahat)= B_\nu(T(\gammahat))$. Once the temperature map has been produced, we can rotate it so that $\betahat = \hat{\bm z}$, and then use the harmonic boost equation to correct for the motion-induced effects (Eq. \eqref{Generalized_aberration_kernel}) and finally rotate it back. Since temperature is not a frequency-dependent parameter, all the frequency derivatives will vanish and only the $n=0$ term will survive and the generalized kernel elements in Eq. \eqref{Generalized_aberration_kernel} simplify to the DC kernel elements with $d=1$

 \begin{equation}\label{Tlm_kernel}
 \tilde{a}^T_{\ell' m} =
 \sum_{\ell}~_0^1\mathcal{K}^{m}_{\ell' \ell}(\beta) ~a^T_{\ell m}.
 \end{equation}
Here  $\tilde{a}^T_{\ell' m} $ are the harmonic multipoles of the temperature observed in the moving frame. The top panel of Fig.~\ref{fig:kernel_m_0} shows the absolute value of $~_0^1\mathcal{K}^{0}_{\ell' \ell}(0.00123)$ for a neighborhood of $\Delta \ell =8$ around different values of $\ell'$. The value of the aberration kernel $~_0^1\mathcal{K}^{m}_{\ell' ,\ell'+\Delta\ell}$ shows how much power from the mode $(\ell',m)$ harmonic mode in temperature leaks into its nearby mode $(\ell'+\Delta\ell,m)$. Since the temperature aberration kernel has the symmetric property $~_0^1\mathcal{K}^{m}_{\ell' ,\ell} = (-1)^{\ell+\ell'}~_0^1\mathcal{K}^{m}_{\ell ,\ell'}$ \cite{Dai2014}, Fig. \ref{fig:kernel_m_0} also shows how much power leaks from the mode $(\ell'+\Delta\ell,m)$ into  $(\ell',m)$. Therefore, it is easy to conclude that if there is equal amount of power in nearby temperature multipoles (e.g. a flat power spectrum), the boost will not have any effect on the observed harmonic modes. Also, here we emphasize again that since $\betahat=\hat{\bm z}$, there will be no leakage from $m'$ into $m\neq m'$ and so the azimuthal modes do not mix under the boost. As we can see in the figure, the leakage of the nearby modes are relatively small for low values of $\ell'$, but the kernel becomes wider as $\ell'$ grows. So, more power leaks from each harmonic mode into its nearby multipoles and vice versa as we go to smaller angular scales. This characteristic of the aberration kernel also emerges when we study the power spectra in a masked sky (\S \ref{sec:IIIC}).

Alternatively, we can choose to employ the integrated intensity $I=\int I_\nu d\nu$ as the observable, instead of the thermodynamic temperature $T$ (e.g. Ref. \cite{Challinor2002}). In this case, since the integrated intensity has a Doppler weight of 4 (see Table \ref{tab:1}), we will need to use the generalized Doppler and aberration kernel with $d=4$ to calculate or correct the effect of boost on the harmonic multipoles. Again, since the integrated intensity is frequency-independent, the harmonic boost equation will simplify to

 \begin{equation}\label{Ilm_kernel}
\tilde{a}^I_{\ell' m} =
\sum_{\ell}~_0^4\mathcal{K}^{m}_{\ell' \ell}(\beta) ~a^I_{\ell m}.
\end{equation}
The middle panel of Fig. \ref{fig:kernel_m_0} shows the absolute relative difference $\Delta \mathcal{K}_I=| ~_0^4\mathcal{K}^{0}_{\ell' \ell}(\beta)/~_0^1\mathcal{K}^{0}_{\ell' \ell}(\beta)|-1$ for different values of $\ell'$. The general behavior of the aberration kernel  $~_0^4\mathcal{K}^{m}_{\ell' \ell}(\beta)$ is similar to that of the thermodynamic temperature $~_0^1\mathcal{K}^{m}_{\ell' \ell}(\beta)$, but its values are typically larger due to the higher Doppler weight of the integrated intensity (see Eq. \eqref{recursive_1}).  The difference is extremely suppressed for $\Delta \ell=0$, but for larger $\Delta \ell$, it is generally more pronounced  at lower $\ell'$. 

Both temperature and integrated intensity can be used as observables for boost correction or detection. The main disadvantage in using these variables is that due to their frequency-independent nature, they do not allow us to examine individual frequency maps, however, this can be alleviated by using the frequency-dependent specific intensity as observable and its corresponding Doppler and aberration kernel.

\subsubsection{Specific Intensity}\label{sec:IIIA2}

 In practice, the CMB observations are performed over narrow frequency bands, with the specific intensity $I_\nu$ as the observable. The generalized Doppler and aberration kernel can be used to apply/correct a boost on individual frequency maps, rather than a combined temperature or integrated intensity map. This is especially advantageous when different masks are applied over different frequencies. For the frequency-dependent observable $I_\nu$, we can write the harmonic boost equation as 

\begin{equation}\label{Inulm_kernel}
\tilde{a}^{I_{\nu}}_{\ell' m}(\nu') =
\sum_{\ell} ~_{0}^{3}\Khat^{m}_{\ell' \ell}(\beta,\nu')  a_{\ell m}^{I_\nu}(\nu).
\end{equation}
Here, since the harmonic multipoles $\tilde{a}^{I_\nu}_{\ell' m} (\nu)$ are frequency-dependent, Eq. \eqref{Inulm_kernel}  does not simplify to a single DC kernel. Instead it will be a linear combination of different Doppler weight DC kernel elements, with coefficients proportional to frequency derivatives of the harmonic multipoles $\tilde{a}^{I_\nu}_{\ell' m} (\nu)$ (Eq. \eqref{Generalized_aberration_kernel}). 

\begin{figure}[t]
	\centering
	\includegraphics[width=1.05\linewidth]{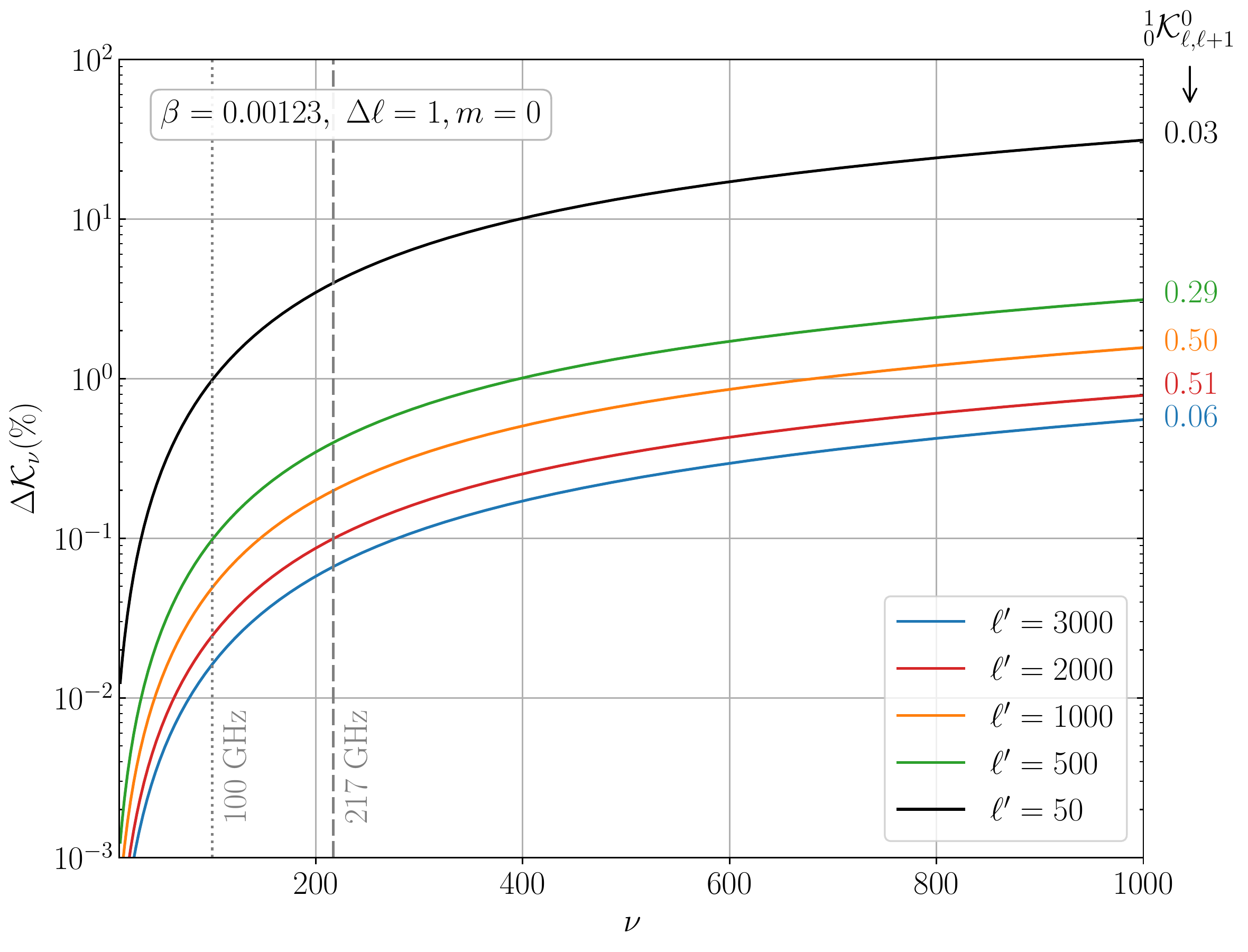}
	\caption{The absolute relative difference between the  aberration kernel for specific intensity $(d=3)$ and thermodynamic temperature $(d=1)$ for $\Delta \ell =1$: $\Delta \mathcal{K}_{I_{\nu}}=| ~_0^3K^{0}_{\ell', \ell' +1}(\beta,\nu')/~_0^1\mathcal{K}^{0}_{\ell' \ell'+1}(\beta)|-1$.  These values show the effective change on $a_{\ell m}^{I_\nu}(\nu)$ over different frequencies. The numbers on the right side show the numerical value of the temperature aberration kernel (frequency-independent) for the corresponding lines with the same color. The deviation of the specific intensity aberration kernel from the temperature kernel is larger at lower $\ell$ values and higher frequencies.  }
	\label{fig:kernel_nu_delta1}
\end{figure}

For a pure blackbody CMB radiation at a temperature $T_0=2.725K$, the frequency dependence of the specific intensity harmonic multipoles is: \cite{Yasini:2016pby} 

\begin{equation}\label{alm_nu=F*alm_T}
a^{I_{\nu}}_{\ell m}(\nu) =F_\nu(T_0) a_{\ell m}^{T}/T_0.
\end{equation}
By plugging this in Eq. \eqref{Inulm_kernel}, and applying the generalized kernel operator on $F_\nu(T_0)$ we obtain 

\begin{align}\label{Inulm_kernel_normalized}
\tilde{a}^{I_{\nu}}_{\ell' m}(\nu') =
&\sum_{\ell} [~_{0}^{3}\Khat^{m}_{\ell' \ell}(\beta,\nu')  F_\nu(T_0)] a_{\ell m}^{T}/T_0\\
\equiv &\sum_{\ell} ~_{0}^{3}K^{m}_{\ell' \ell}(\beta,\nu')  a_{\ell m}^{I_\nu}(\nu').
\end{align}
where the normalized kernel $~_{s}^{d}K^{m}_{\ell' \ell}(\beta,\nu') $ defined in Eq. \eqref{normalized_kernel},
can be thought of as the eigenfunction of the operator $~_{s}^{d}\Khat^{m}_{\ell' \ell}(\beta,\nu')$. The bottom panel of Fig \ref{fig:kernel_m_0} shows the absolute relative difference between the frequency-dependent normalized aberration kernel elements ($d=3$) and the DC temperature kernel ($d=1$) which we represent as $\Delta \mathcal{K}_{I_{\nu}}=| ~_0^3K^{0}_{\ell' \ell}(\beta,\nu')/~_0^1\mathcal{K}^{0}_{\ell' \ell}(\beta)|-1$. The behavior is similar to the aberration kernel for the integrated intensity, but since the specific intensity has a lower Doppler weight, its aberration kernel is overall smaller. Similarly, since the specific intensity has a higher Doppler weight than the thermodynamic temperature, its kernel elements are also larger than the DC aberration kernel elements with $d=1$. 

The most important distinction between the normalized aberration kernel for specific intensity and the DC kernels for temperature and integrated intensity is the frequency dependence. In Fig. \ref{fig:kernel_m_0}, we have shown $\Delta \mathcal{K}_{I_{\nu}}$ at two different frequencies of 100 GHz and 217 GHz. These representative values  correspond to two of the frequency bands used by the \emph{Planck} collaboration for cosmological parameter estimation \cite{Aghanim:2015xee}. The values of the normalized aberration kernel elements $~_0^3K^{0}_{\ell' \ell}(\beta,\nu')$ are smaller at the lower frequency of 100 GHz. In fact as $\nu' \rightarrow 0$ we get $~_0^3K^{0}_{\ell' \ell}(\beta,\nu') \rightarrow ~_0^3\mathcal{K}^{0}_{\ell' \ell}(\beta) $ and therefore the specific intensity aberration kernel becomes smaller and converges to its $d$-equivalent frequency-independent DC aberration kernel in the Rayleigh-Jeans limit. This convergence happens for any radiation with a power law frequency spectrum $\nu^\alpha$ ($\alpha =2$ for a blackbody at low frequencies). After applying the frequency derivatives $\nu'^n \partial_{\nu'}$ (see Eq. \eqref{Generalized_aberration_kernel}) on a power-law spectrum we obtain $\nu'^n \alpha! / (\alpha-n)!$; at $\nu' =0$ this expression is only non-zero for $n=0$ (The DC term), so in the expression for the generalized aberration kernel, all the higher order correction due to the frequency dependence of the observable vanish, and the kernel converges to its DC equivalent with the same Doppler weight. Fig. \ref{fig:kernel_nu_delta1} shows the frequency dependence of the specific frequency aberration kernel only for the first neighbor $\Delta\ell=1$, using the parameter $\Delta \mathcal{K}_{{\nu}}\equiv |~_0^3K^{0}_{\ell',\ell'+1}(\beta,\nu')/~_0^1\mathcal{K}^{0}_{ \ell',\ell'+1}(\beta)|-1$ for different values of $\ell'$. This is essentially an extension of the bottom panel of Fig. \ref{fig:kernel_m_0} for $\Delta\ell=1$ over different frequencies. The relative difference at $\nu'=0$ shows the offset between $~_0^3\mathcal{K}^{0}_{\ell' \ell}(\beta) $ and $~_0^1\mathcal{K}^{0}_{\ell' \ell}(\beta) $ for different values of $\ell'$. Since the Doppler weight of the kernel becomes irrelevant at smaller angular scales, the offset becomes smaller as $\ell'$ grows. At higher frequencies, however, the relative difference between the aberration kernels for specific intensity and thermodynamic temperature grows and the frequency dependence of the observable becomes non-negligible. The relative difference can exceed $10\%$ for low $\ell'$ modes (e.g. $\ell'=50$ at 400 GHz); however, one should note that the value of the temperature kernel itself (reported on the right hand side of the plot) becomes smaller for these modes. 

As we can see in the bottom panel of Fig. \ref{fig:kernel_m_0} and Fig. \ref{fig:kernel_nu_delta1}, the modulus of the aberration kernel elements (which represent the motion-induced correlation between nearby multipoles) are consistently larger for $~_0^3K^{0}_{\ell' \ell}(\beta)$ elements than $~_0^1\mathcal{K}^{0}_{\ell' \ell}(\beta)$. Therefore, employing $I_\nu$ instead of $T$ would be beneficial in reconstruction of $\vec{\bm \beta}$  using the multipole correlation (outlined in \cite{Aghanim:2013suk}) because it will lead to a stronger motion-induced mode coupling and therefore higher signal to noise for the effect.

\subsubsection{Polarized Temperature and Integrated Intensity}\label{sec:IIA3}

In this section we turn our focus to the motion induced effects on the polarization of the  CMB. The harmonic boost equation for E and B polarization modes can be written in terms of the DC kernel elements with $s = \pm 2$ \cite{Dai2014,Challinor2002,Notari:2011sb}
\begin{subequations}\label{ebT_harmonic_boost}
\begin{align}
\tilde{e}^T_{\ell' m} &= \sum_{\ell} (_\text{ee}^{~1}\mathcal{K}^m_{\ell' \ell}(\beta) e^T_{\ell m} +  _\text{eb}^{~1}\mathcal{K}^m_{\ell' \ell}(\beta) b^T_{\ell m}),\\
\tilde{b}^T_{\ell' m} &= \sum_{\ell}(_\text{be}^{~1}\mathcal{K}^m_{\ell' \ell}(\beta) e^T_{\ell m} +  _\text{bb}^{~1}\mathcal{K}^m_{\ell' \ell}(\beta) b^T_{\ell m}),
\end{align}
\end{subequations}
where 
\begin{subequations}\label{polarized_kernel}
\begin{align}
_\text{ee}^{~d}\mathcal{K}^m_{\ell' \ell}(\beta) &\equiv \frac{1}{2}[~_{2}^{d}\mathcal{K}^m_{\ell' \ell}(\beta) + _{-2}^{~~d}\mathcal{K}^m_{\ell' \ell}(\beta)~ ], \\
_\text{eb}^{~d}\mathcal{K}^m_{\ell' \ell}(\beta)&\equiv \frac{i}{2}[~_{2}^{d}\mathcal{K}^m_{\ell' \ell}(\beta) - _{-2}^{~~d}\mathcal{K}^m_{\ell' \ell}(\beta)~ ], \\
_\text{be}^{~d}\mathcal{K}^m_{\ell' \ell}(\beta) &\equiv- _\text{eb}^{~d}\mathcal{K}^m_{\ell' \ell}(\beta),  \\
_\text{bb}^{~d}\mathcal{K}^m_{\ell' \ell}(\beta)& \equiv~ _\text{ee}^{~d}\mathcal{K}^m_{\ell' \ell}(\beta). 
\end{align}
\end{subequations}
The kernel coefficients $_\text{ee}^{~d}\mathcal{K}^m_{\ell' \ell}(\beta)$ and $_\text{bb}^{~d}\mathcal{K}^m_{\ell' \ell}(\beta)$ represent the boost modifications to the E and B modes (auto-leakage), while $_\text{eb}^{~d}\mathcal{K}^m_{\ell' \ell}(\beta)$ and $_\text{be}^{~d}\mathcal{K}^m_{\ell' \ell}(\beta)$ respectively capture the motion-induced E to B leakage and vice versa (cross-leakage).

\begin{figure}[t]
	\centering
	\includegraphics[width=1.0\linewidth]{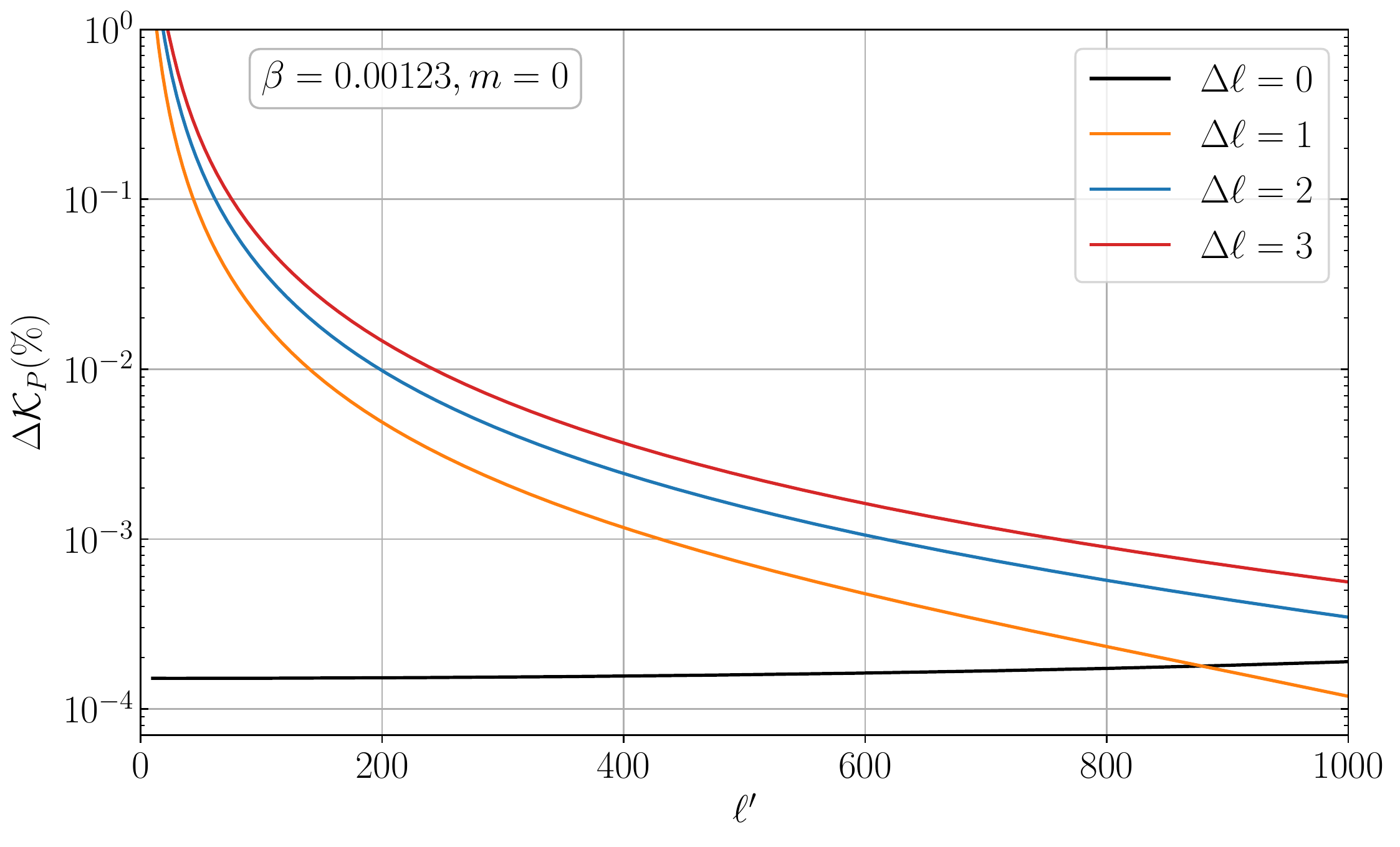}
	\caption{\emph{top}: The absolute relative difference between the aberration kernel elements for polarized and unpolarized thermodynamic temperature over different angular modes. The difference becomes smaller as $\ell'$ grows, except for the central value of the kernel ($\Delta \ell =0$) which is almost spin ($s$) independent.}
	\label{fig:delta_kernel_p}
\end{figure}

It is easy to calculate the DC aberration kernel for polarization $~_{\pm2}^{~~1}\mathcal{K}^m_{\ell' \ell}(\beta) $ using the kernel ODE (Eq. \eqref{ODE}). The main difference between this case and the temperature aberration kernel $(s=0)$, originates in the $_s\C_{\ell'm} $ coefficients. As it is evident from Eq. \eqref{Clm}, the difference between $_2\C_{\ell'm} $ and $_0\C_{\ell'm} $ becomes smaller as $\ell'$ grows. Therefore, the difference between the DC polarization and temperature aberration kernels become negligible for large values of $\ell'$. Fig. \ref{fig:delta_kernel_p} shows this convergence between the aberration kernels with different spin weights, using the relative difference $\Delta \mathcal{K}_P \equiv |~_{2}^{1}\mathcal{K}^{0}_{\ell' \ell}(\beta) /_{0}^{1}\mathcal{K}^{0}_{\ell' \ell}(\beta)|-1$ for different values of $\Delta\ell = \ell - \ell'$ and $m=0$. As it is evident from the plot, the difference is only greater than $0.1\%$ for $\ell' \lesssim 100$, while at smaller scales $~_{2}^{1}\mathcal{K}^{0}_{\ell' \ell}(\beta)$ converges to the DC temperature kernel $~_{0}^{1}\mathcal{K}^{0}_{\ell' \ell}(\beta)$. The difference is indeed even smaller for $m>0$, so at large values of $\ell'$ one can safely use the unpolarized aberration kernel for boost corrections of the polarized observables. However, in what follows, for the sake of accuracy, we do not make this simplification and use the actual values of the polarization kernel obtained from solving the kernel ODE.

Since the $_s\C_{\ell' m} $ coefficients are not sensitive to the sign of $s$, it is easy to see  from the kernel ODE (Eq. \eqref{ODE}) that $_{2}^{1}\mathcal{K}^{m}_{\ell' \ell} =~ _{-2}^{~1}\mathcal{K}^{m}_{\ell' \ell}$. This allows us to simplify the polarization aberration kernels in Eqs. \eqref{polarized_kernel} as 

\begin{subequations}\label{simp_polarized_kernel}
	\begin{align}
	_\text{ee}^{~1}\mathcal{K}^m_{\ell' \ell}(\beta) &=~	_\text{bb}^{~1}\mathcal{K}^m_{\ell' \ell}(\beta) = ~_{2}^{1}\mathcal{K}^m_{\ell' \ell}(\beta),\\
	_\text{eb}^{~1}\mathcal{K}^m_{\ell' \ell}(\beta)&=~_\text{be}^{~1}\mathcal{K}^m_{\ell' \ell}(\beta)= 0.
	\end{align}
\end{subequations}
Therefore, for polarized temperature there is no E to B leakage (and vice versa). As it is shown in Ref. \cite{Dai2014}, this condition only holds for $d=1$ and in general for observables with Doppler weights other than 1, $_{2}^{d}\mathcal{K}^m_{\ell' \ell}(\beta) \neq _{-2}^{~~d}\mathcal{K}^m_{\ell' \ell}(\beta) $. Any difference between the kernel elements with $s=2$ and $s= -2$ will lead to a spurious E to B leakage. For example, when using integrated intensity $(d=4)$, there will be an E to B leakage that becomes as large as $\sim 3 \times 10^{-4}\mu$K at $\ell'=2$ \cite{Challinor2002}. 

The polarization kernel elements for integrated intensity have been studied in detail in Ref. \cite{Challinor2002}, so we do not discuss them here any further. Instead, we turn our attention to the frequency-dependent polarized specific intensity. 

\begin{figure}[t]
	\centering
	\includegraphics[width=1.0\linewidth]{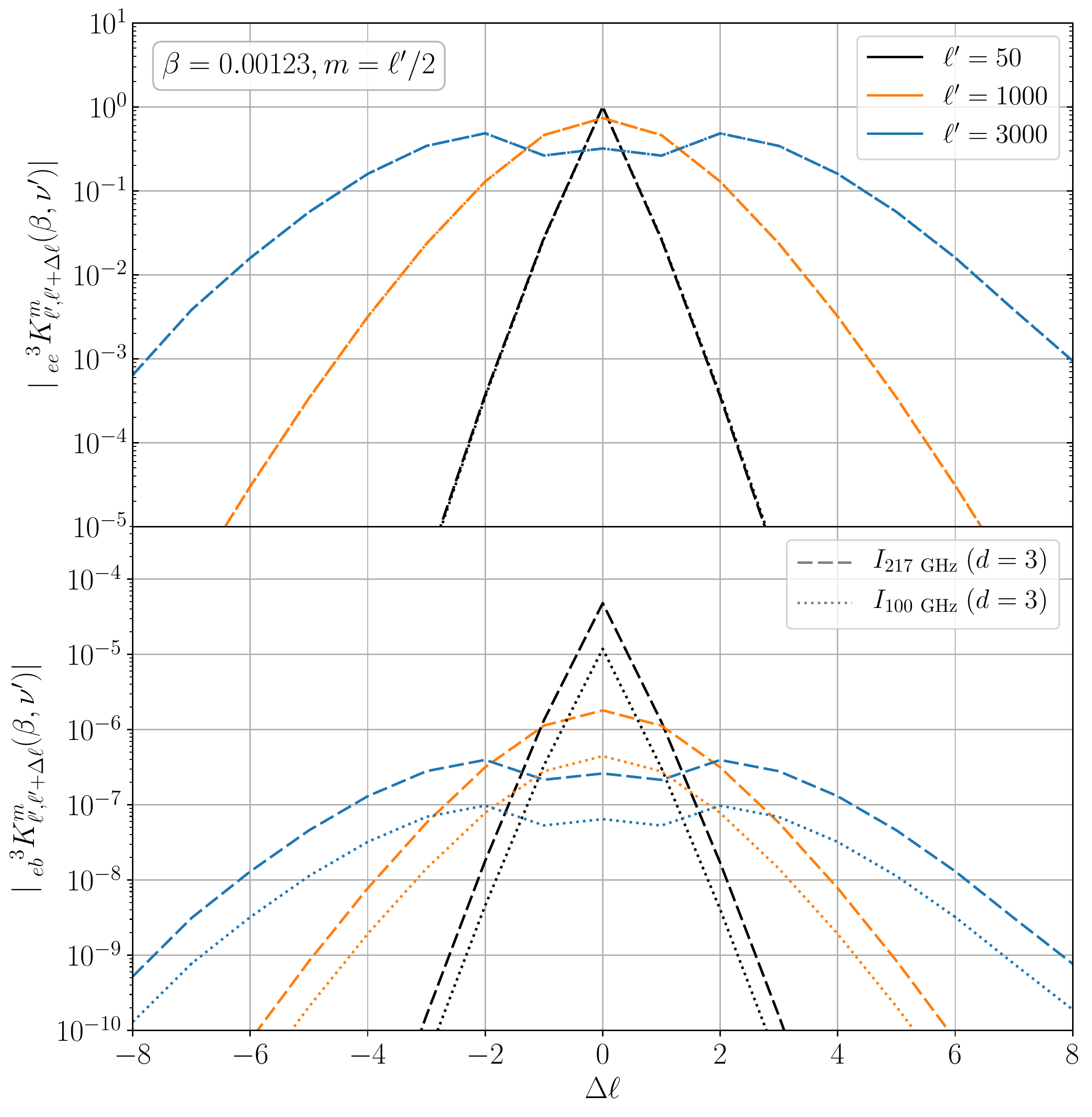}
	\caption{The normalized aberration kernel elements for polarized specific intensity at 100 and 217 GHz  \emph{top}: The modulus of the kernel for $EE$ and $BB$ modes. The values of the kernel elements are almost identical to the unpolarized case at high $\ell'$. The difference between different frequencies are not visible in the plot. \emph{bottom}: The modulus of the kernel for $EB$ and $BE$ leakage. The values are smaller at the lower frequency of 100 GHz compared with 217 GHz and also negligible compared with the auto-leakage components ($EE$ and $BB$).}
	\label{fig:kernel_Inu_ee_eb}
\end{figure}

\subsubsection{Polarized Specific Intensity}
Using the generalized Doppler and aberration kernel operator we can write the harmonic boost equation for polarized specific intensity as 

\begin{subequations}\label{ebI_nu_harmonic_boost}
	\begin{align}
	\tilde{e}^{I_\nu}_{\ell' m}(\nu') &= \sum_{\ell}~_\text{ee}^{~3}K^m_{\ell' \ell}(\beta,\nu') e^{I_\nu}_{\ell m}(\nu') +  _\text{eb}^{~3}K^m_{\ell' \ell}(\beta,\nu') b^{I_\nu}_{\ell m}(\nu'),\\
	\tilde{b}^{I_\nu}_{\ell' m} (\nu')&= \sum_{\ell}~_\text{be}^{~3}K^m_{\ell' \ell}(\beta,\nu') e^{I_\nu}_{\ell m}(\nu') +  _\text{bb}^{~3}K^m_{\ell' \ell}(\beta,\nu') b^{I_\nu}_{\ell m}(\nu'),
	\end{align}
\end{subequations}
where similar to Eqs. \eqref{polarized_kernel}, we define the polarized specific intensity aberration kernel elements as 
\begin{subequations}\label{polarized_kernel_nu}
	\begin{align}
	_\text{ee}^{~3}K^m_{\ell' \ell}(\beta,\nu') &\equiv \frac{1}{2}[~_{2}^{3}K^m_{\ell' \ell}(\beta,\nu') + _{-2}^{~~3}K^m_{\ell' \ell}(\beta,\nu')~ ], \\
	_\text{eb}^{~3}K^m_{\ell' \ell}(\beta,\nu')&\equiv \frac{i}{2}[~_{2}^{3}K^m_{\ell' \ell}(\beta,\nu') - _{-2}^{~~3}K^m_{\ell' \ell}(\beta,\nu')~ ], \\
	_\text{be}^{~3}K^m_{\ell' \ell}(\beta,\nu') &\equiv- _\text{eb}^{~3}K^m_{\ell' \ell}(\beta,\nu'),  \\
	_\text{bb}^{~3}K^m_{\ell' \ell}(\beta,\nu')& \equiv~ _\text{ee}^{~3}K^m_{\ell' \ell}(\beta,\nu'). 
	\end{align}
\end{subequations}

We start with the interesting case of E to B aberration kernel. The bottom panel of Fig. \ref{fig:kernel_Inu_ee_eb} shows the modulus of the $_\text{eb}^{~3}K^{\ell'/2}_{\ell' \ell}(\beta,\nu')$ and equivalently $_\text{be}^{~3}K^{\ell'/2}_{\ell' \ell}(\beta,\nu')$ at two different frequencies of 100 GHz and 217 GHz. Since according to recursive Eqs. \eqref{recursive_1}, the two kernel elements $_\text{+2}^{~3}K^{m}_{\ell' \ell}(\beta,\nu')$ and $_\text{-2}^{~3}K^{m}_{\ell' \ell}(\beta,\nu')$ are identical for $m=0$, the value of $_\text{eb}^{~3}K^{m=0}_{\ell' \ell}(\beta,\nu')$ is always zero for this mode. Therefore, we plot the kernel elements for the arbitrary value of $m=\ell'/2$. As we see in the plot, the aberration kernel $_\text{eb}^{~3}K^{\ell'/2}_{\ell' \ell}(\beta,\nu')$ is relatively small even at low $\ell'$, so we do not expect the motion-induced E to B polarization leakage to be a major source of error in CMB B mode observations (we will study this case further in the section dedicated to polarization power spectrum \S \ref{sec:IIIB4}).

Even though the cross-component leakage in polarization seems to be negligible, the motion-induced modifications to each individual mode cannot be easily neglected. Similar to the unpolarized case,  the E mode and B mode polarization of the CMB will be affected by the boost. The top panel of Fig. \ref{fig:kernel_Inu_ee_eb} shows the modulus of the $_\text{ee}^{~3}K^{\ell'/2}_{\ell' \ell}(\beta,\nu')$ and equivalently $_\text{bb}^{~3}K^{\ell'/2}_{\ell' \ell}(\beta,\nu')$ at two different frequencies of 100 GHz and 217 GHz. The behavior of the polarization aberration kernel is very similar to the unpolarized ones, in that it is sharper at low values of $\ell'$, but becomes wider as $\ell'$ grows. The difference between the two frequencies is of the order of $\sim 10^{-3}$ and therefore not visible in the plot. This difference, however, will prove important when we study the effect of boost on masked skies. 
 	  
\subsection{All-sky Power Spectra}\label{sec:IIIB}
So far we have examined individual kernel elements and their variation with Doppler weight and frequency. However, what we are ultimately interested in is the motion-induced effects on the power spectrum. The statistical effect of boost on the power spectrum can be simplified as 

\begin{equation}
_s^d\tilde{C}_{\ell' }= \sum_{\ell}~_s^d\mathcal{B}_{\ell' \ell}(\beta)~_s^dC_\ell,
\end{equation}
where 
\begin{equation}
_s^dC_\ell = \frac{1}{2\ell+1} \sum_m \langle |_s^dX_{\ell m}|^2 \rangle
\end{equation} 
is the power spectrum of the observable $ _s^dX_{\ell m}$ in the CMB frame and $_s^d\tilde{C}_{\ell' }$ is the boosted power spectrum. Here we have acquired the notion of the \emph{boost power transfer matrix} (BPTM) defined in Ref. \cite{Jeong:2013sxy} as

\begin{equation}\label{BPTM_d}
_s^d\mathcal{B}_{\ell' \ell}(\beta) \equiv \frac{1}{2\ell' +1} \sum_{m} \Big| ~_s^d\mathcal{K}^m_{\ell' \ell}(\beta)~\Big|^2.
\end{equation}
We use a parallel definition of the boost power transfer matrix for the normalized frequency-dependent aberration kernel elements as
\begin{equation}\label{BPTM_nu}
_s^dB_{\ell' \ell}(\beta,\nu') \equiv \frac{1}{2\ell' +1} \sum_{m} \Big| ~_s^dK^m_{\ell' \ell}(\beta,\nu')~\Big|^2.
\end{equation}
The theoretical advantage of using BPTM for studying the motion-induced effects is that it can be directly applied to analytic power spectra and it is not prone to random map realization noise and cosmic variance. In other words, it provides information about how the boost affects the statistical ensemble of the harmonic multipoles. In this subsection we study the effects of boost on the power spectrum and its dependence on the Doppler weight and frequency of observation, first by analyzing the BPTM and then by looking at simulated CMB skies.

\subsubsection{Thermodynamic Temperature and Integrated Intensity Power Transfer}

\begin{figure}[t]
	\centering
	All-sky
	\includegraphics[width=1.0\linewidth]{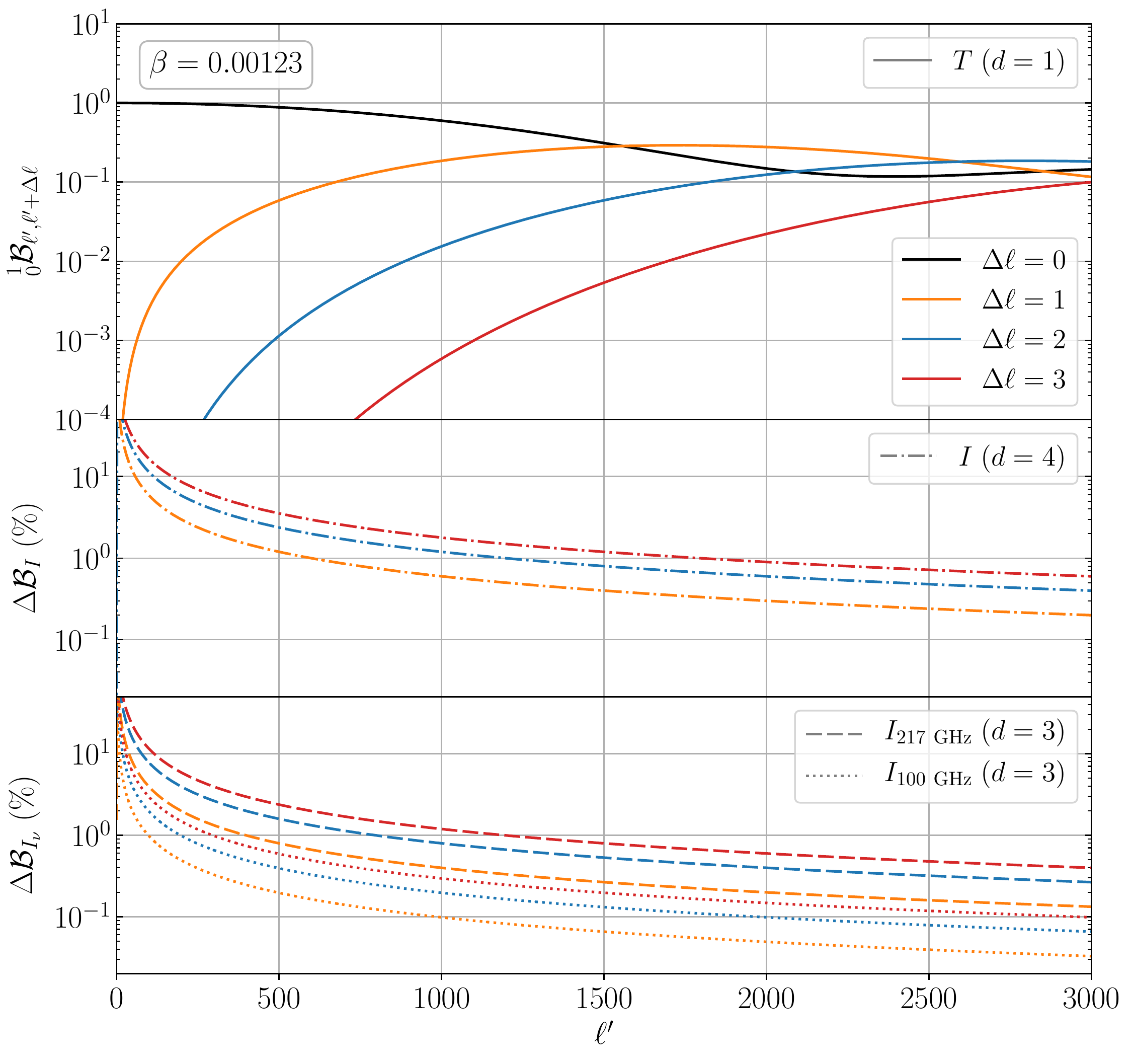}
	\caption{\emph{top}: The \emph{ Boost Power Transfer Matrix} for temperature aberration kernel. There is no significant power transfer at $\ell'<800$, but at smaller angular scales more power leaks from $\Delta \ell=0$ to the nearby neighbors. \emph{middle}: The absolute relative difference between the BPTM for integrated intensity and temperature $\Delta \mathcal{B}_{I} \equiv |_0^4\mathcal{B}_{\ell' \ell}(\beta)/_0^1\mathcal{B}_{\ell' \ell}(\beta)|-1$ for $\ell = \ell' + \Delta \ell$. The relative difference is larger than 1\% at low $\ell'$ ($\lesssim 500$) , but the value of the aberration kernel is also negligibly small over these angular scales. \emph{bottom}: The absolute relative difference between the BPTM for integrated intensity and temperature $\Delta \mathcal{B}_{I} \equiv |_0^3B_{\ell' \ell}(\beta,\nu')/_0^1\mathcal{B}_{\ell' \ell}(\beta)|-1$. The BPTM for specific intensity is smaller than Integrated intensity over all angular scales, and its relative difference with the temperature BPTM decreases at lower frequencies.   }
	\label{fig:boost_power_transfer_unmasked}
\end{figure}

First, let us start by looking at the boost power transfer matrix for the DC temperature kernel $_0^1\mathcal{K}^m_{\ell' \ell'+\Delta\ell}$. As it can be seen in Fig. \ref{fig:kernel_m_0}, the central value of the aberration kernel ($\Delta\ell=0$) typically gets smaller as $\ell'$ grows. This can be interpreted as transfer of power from $\Delta\ell=0$ to the nearby neighbors $\Delta\ell=1, 2,3 $ etc., which grow larger at higher $\ell'$s. Naturally we would expect the same behavior from the BPTM of the temperature kernel $_0^1\mathcal{B}_{\ell',\ell'+\Delta \ell}(\beta)$, which shows how much power leaks to the nearby multipoles $\Delta \ell$ at each observed angular mode $\ell'$. The top panel of Fig.  \ref{fig:boost_power_transfer_unmasked} shows $_0^1\mathcal{B}_{\ell',\ell'+\Delta \ell}(0.00123)$ for $\ell'_\text{max}=3000$. As it is expected from the general behavior of the aberration kernel, at large angular scales (small $\ell'$) most of the power remains within the same $\ell'$ mode ($\Delta \ell=0$), but it starts to leak into the nearby multipoles as $\ell'$ increases. This can be simply explained by considering the angle change due to aberration effect which is of the order $\Delta \theta \simeq \beta |\betahat \times \gammahat|$. For the dipole-inferred value of $\beta$, this would correspond to $\Delta \theta_\text{max} \simeq 0.00123$. For $\ell' < 1/\Delta \theta_\text{max} \simeq 800$, since the angular change due to motion is smaller than the angular scale of the harmonic multipoles, the aberrated photons remain within the same angular modes. Therefore, no significant power leakage to nearby multipoles is expected at $\ell' \lesssim 800$. As the aberration kernel becomes wider at larger $\ell'$s, more power leaks into further neighbors. For example, the power leakage into $\Delta \ell =1$ becomes larger than $\Delta \ell =0$ at $\ell'\simeq1500$, and roughly at $\ell' \simeq 2600$ the  leakage into $\Delta \ell =2$ becomes dominant over $\Delta \ell =1$ and so on. 

As mentioned earlier, the value of the DC aberration kernel elements for integrated intensity with $d=4$ are typically larger than the corresponding elements for temperature with $d=1$ (see Fig. \ref{fig:kernel_m_0}). Consequently, the power transfer matrix $_0^4\mathcal{B}_{\ell' \ell}(\beta)$ is expected to be larger than $_0^1\mathcal{B}_{\ell' \ell}(\beta)$ over all angular scales. The middle panel of Fig. \ref{fig:boost_power_transfer_unmasked} shows the percent difference between the two BPTMs: $\Delta \mathcal{B}_{I} \equiv |_0^4\mathcal{B}_{\ell' \ell}(\beta)/_0^1\mathcal{B}_{\ell' \ell}(\beta)|-1$ for $\ell = \ell' + \Delta \ell$ and $\beta=0.00123$. At large angular scales, the relative difference between the BPTMs for $\Delta \ell >0$ is large, mainly due to the fact that the value of the temperature BPTM (top panel of Fig.\ref{fig:boost_power_transfer_unmasked}) is small. As $\ell'$ grows larger, $\Delta \mathcal{B}_{I}$ becomes smaller and the Doppler weight of the observable becomes irrelevant. Also, as expected from Fig. \ref{fig:kernel_m_0}, the central value of the aberration kernel ($\Delta \ell=0$) is nearly $d$-independent. The relative difference between the BPTMs for $\Delta \ell=0$ has not been shown in the plot, but it does not become larger than $4 \times 10^{-4}$.

\subsubsection{Specific Intensity Power Transfer}
Similar to the integrated intensity, the specific intensity boost power transfer matrix has generally larger values compared with the temperature one. This is simply due to the fact that specific intensity has a larger Doppler weight ($d=3$) than the thermodynamic temperature ($d=1$). The bottom panel of Fig. \ref{fig:boost_power_transfer_unmasked} shows the relative difference between the frequency-dependent power transfer matrix for specific intensity and the temperature power transfer $\Delta \mathcal{B}_{I_\nu} \equiv |_0^3B_{\ell' \ell}(\beta,\nu')/_0^1\mathcal{B}_{\ell' \ell}(\beta)|-1$ at two different frequencies of 100 GHz and 217 GHz. The behavior of $_0^3B_{\ell' \ell}(\beta,\nu')$ for these chosen frequencies closely resemble the BPTM for integrated intensity with the  difference that $_0^3B_{\ell' \ell}(\beta,\nu')$ are smaller over all angular scales due to their smaller Doppler weight $d=3$. For the same reason, the specific intensity BPTM is larger than the temperature BPTM with $d=1$. Also, $\Delta \mathcal{B}_{I_{\text{217GHz}}}$ is larger than $\Delta \mathcal{B}_{I_{\text{100GHz}}}$ by roughly a factor of 4 over all angular scales. For the all-sky maps the difference between the BPTM specific intensity becomes smaller compared with the temperature BPTM.

\begin{figure}[t]
	\centering
	All-sky
	\includegraphics[width=1.0\linewidth]{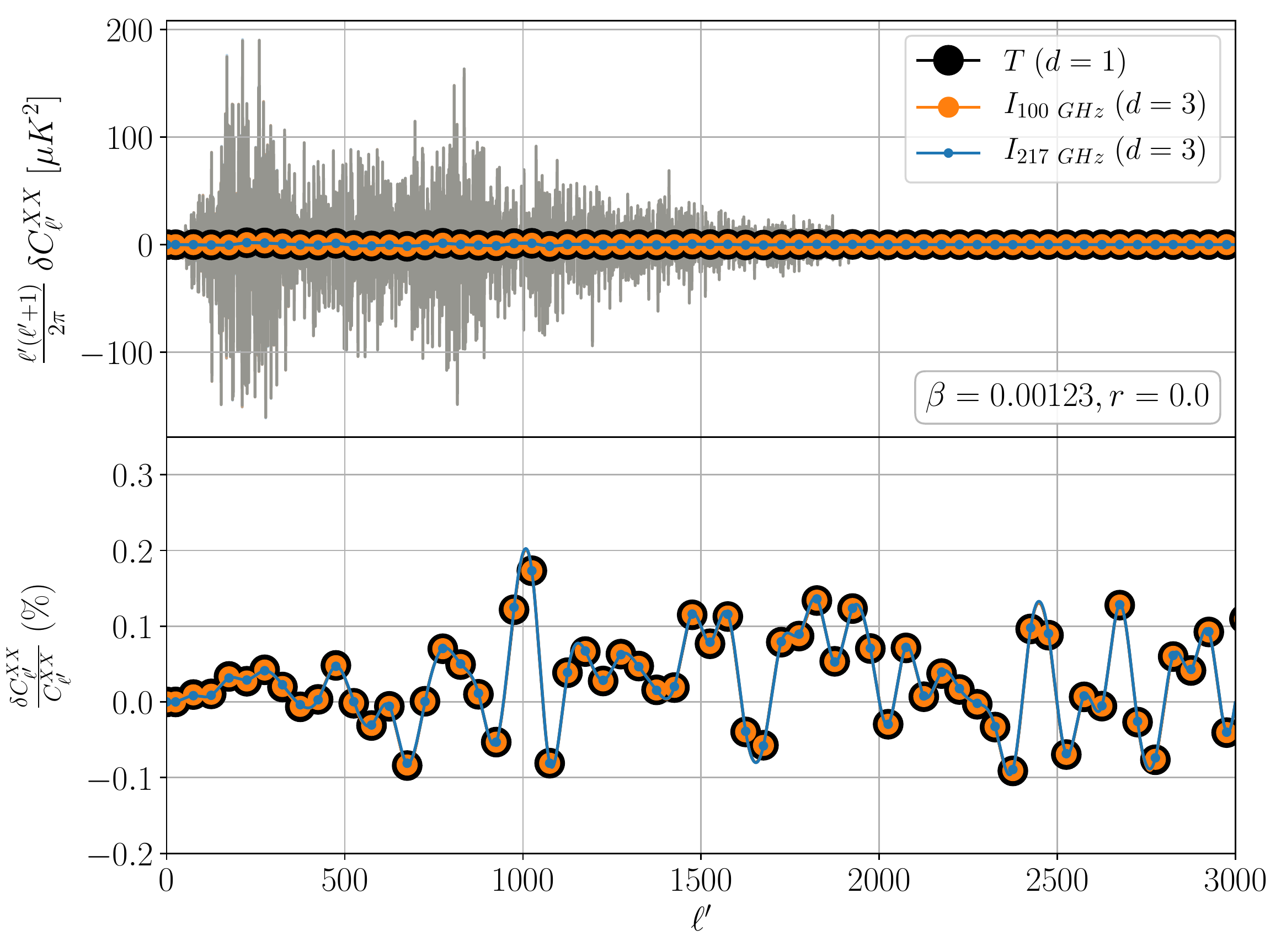}
	\caption{\emph{top}: The difference between boosted and rest frame power spectra for thermodynamic temperature (\emph{black}), specific intensity at 100 GHz (\emph{orange}) and  217 GHz (\emph{blue}) for a single sky realization. The vertical lines (\emph{gray}) show the difference for individual $\ell'$ modes, but the circles are the binned values for $\Delta \ell' =50$ and are interpolated with a cubic spline. \emph{bottom}: The relative difference between boosted and rest frame power spectra shown in the top panel. The fluctuations are at most 0.2\% and relatively flat around 0 over all angular scales. There is no significant difference between the boost in temperature and specific intensity at different frequencies.}
	\label{fig:TT_power_spectrum_unmasked}
\end{figure}

\begin{figure}[t]
	\centering
	All-sky
	\includegraphics[width=1.0\linewidth]{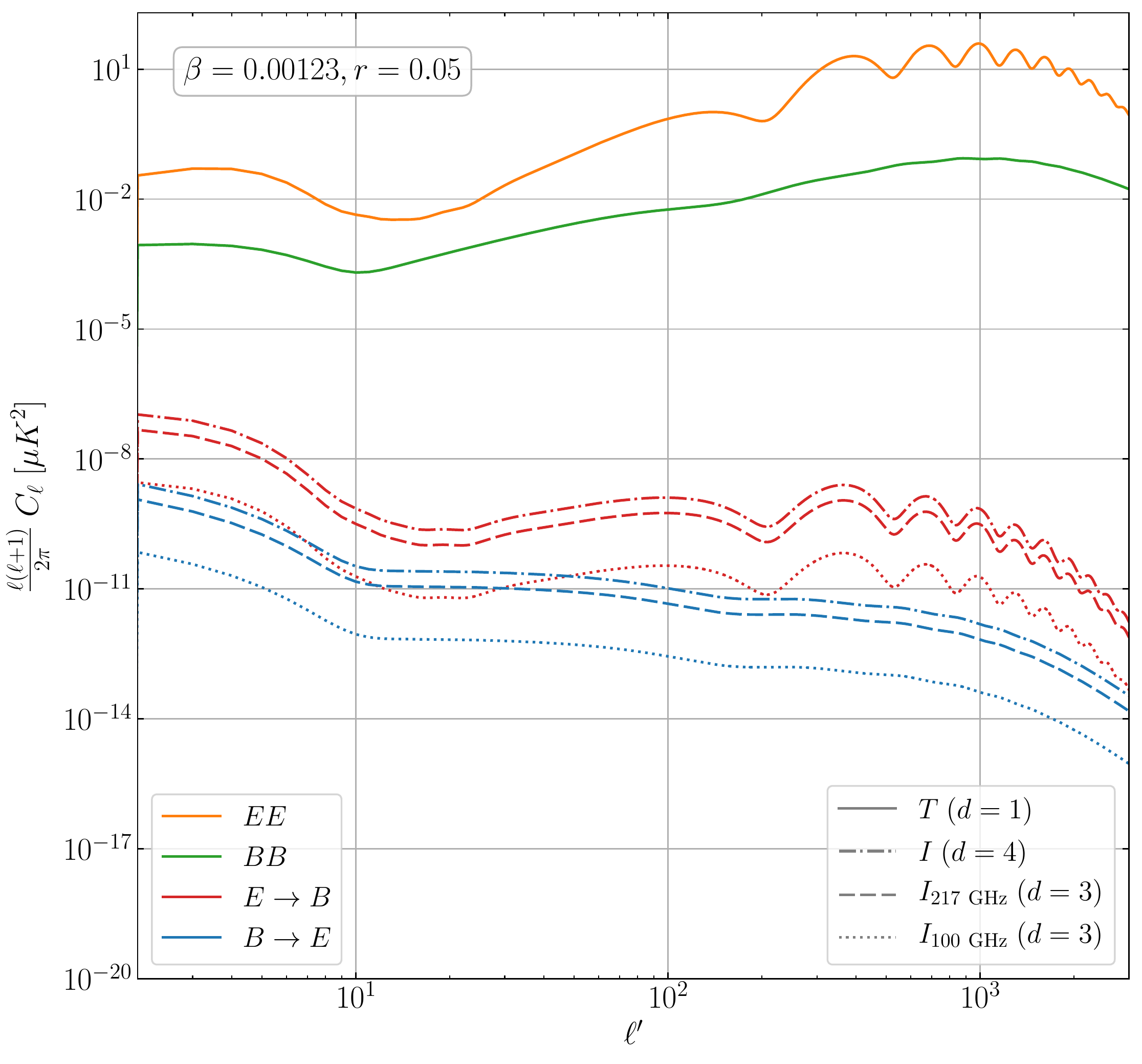}
	\caption{All-sky polarized power spectra in a boosted frame. The boosted $EE$ (\emph{orange}) and $BB$ (\emph{green}) auto-spectra for all the observables lie almost on top of the rest frame lines and therefore are not distinguishable from them. The motion-induced cross-spectra $E$ to $B$ (\emph{red}) and $B$ to $E$ (\emph{blue}) for specific intensity are larger at the higher frequency of 217 GHz compared with 100 GHz, but they are both smaller than the integrated intensity motion-induced cross-spectra. The intensity parameters have been normalized to temperature and the units are converted to $\mu$K. }
	\label{fig:polarized_power_spectrum_unmasked}
\end{figure}

\subsubsection{Unpolarized Power Spectrum}\label{sec:IIIB3}

We showed that the the BPTM for the integrated intensity and specific intensity are typically larger than the one for thermodynamic temperature. The relative difference is larger over larger angular scales (low $\ell'$) where the aberration kernel itself is small. Over smaller angular scales (high $\ell'$) where the kernel values become significantly larger, the relative difference between BPTMs with different Doppler weights becomes negligible. The combination of these variations leads to a relatively constant difference between the BPTMs with different Doppler weights. For an all-sky map, however, this difference is completely negligible as it is depicted in Fig. \ref{fig:TT_power_spectrum_unmasked}. The top panel shows the difference $\delta C^{XX}_{\ell'} = \tilde{C}^{XX}_{\ell'}-C^{XX}_{\ell'}$ between the boosted and rest frame power spectra for a single realization. Here we are using the notation $C^{XX}_\ell$ to represent the power spectrum estimator applied to the maps, where $X$ stands for $T$, $I_{100~\text{GHz}}$ and $I_{217~\text{GHz}}$. The bottom panel shows the relative difference between the two power spectra $\delta C^{XX}_{\ell'}/C^{XX}_{\ell'}$. We calculate $\delta C^{XX}_{\ell'}$ for temperature and using the DC kernel elements $_0^1\mathcal{K}^m_{\ell'\ell}(\beta)$, and for the specific intensity using the normalized frequency-dependent kernel $_0^3K^m_{\ell'\ell}(\beta,\nu')$. The effect of the boost on a single realization can be as large as $0.2 \%$, when binned over $\Delta \ell' = 50$ \footnote{The size of the bin has been set to 50 to allow for easy comparison of our results with similar studies. \cite{Catena:2012hq,Jeong:2013sxy}}, but there is no noticeable difference between the boost in temperature or in specific intensity. The motion-induced effects on the power spectra are generally suppressed in all-sky power spectra due to cancellations between the effect in the antipodal directions of motion, however, as we will show this is not the case for masked skies.

%

\subsubsection{Polarized Power Spectra}\label{sec:IIIB4}

The BPTM and the power spectra for polarization  auto-leakage ($E$ to $E$ and $B$ to $B$) components are almost identical to the ones in the unpolarized case (Fig.~\ref{fig:boost_power_transfer_unmasked}) and therefore we will not discuss them here. Instead, we examine the polarization cross-leakage ($E$ to $B$ and $B$ to $E$) components in an all-sky experiment. Fig.~\ref{fig:polarized_power_spectrum_unmasked} shows all the polarization power spectra in a boosted frame. We adopt a theoretical power spectrum with an arbitrary value of $r=0.05$ generated with CAMB  in the rest frame, and boost it using the BPTMs defined in Eqs. \eqref{BPTM_d} and \eqref{BPTM_nu}. For the $EE$ and $BB$ components, the boosted and rest frame lines are not distinguishable in the plot. The relative difference due to the boost for both these components ($\delta C^{EE}_{\ell'}/C^{EE}_{\ell'}$ and $\delta C^{BB}_{\ell'}/C^{BB}_{\ell'}$) in an all-sky map is almost identical to the one in Fig.~\ref{fig:TT_power_spectrum_unmasked}: they are relatively flat over all angular scales with a maximum fluctuation of 0.2\% for $\Delta \ell' =50$. 

As we discussed in \S\ref{sec:IIA3}, the cross-leakage aberration kernel is zero for polarized temperature, and so is its BPTM. Therefore, we do not expect any $E$ to $B$ or $B$ to $E$ leakage due to boost in polarized temperature $(d=1)$.  However, as mentioned earlier, this is not the case for observables with $d\neq1$. In Fig. \ref{fig:polarized_power_spectrum_unmasked} we show these cross-leakage components for integrated intensity and specific intensity at 100 and 217 GHz. Even though these motion-induced leakage components are non-zero, there are still several orders of magnitude below the $EE$ and $BB$ auto-spectra in an all-sky analysis and likely below both instrumental sensitivity and foreground residuals.

\begin{figure}[t]
	\centering
	\includegraphics[width=0.8\linewidth]{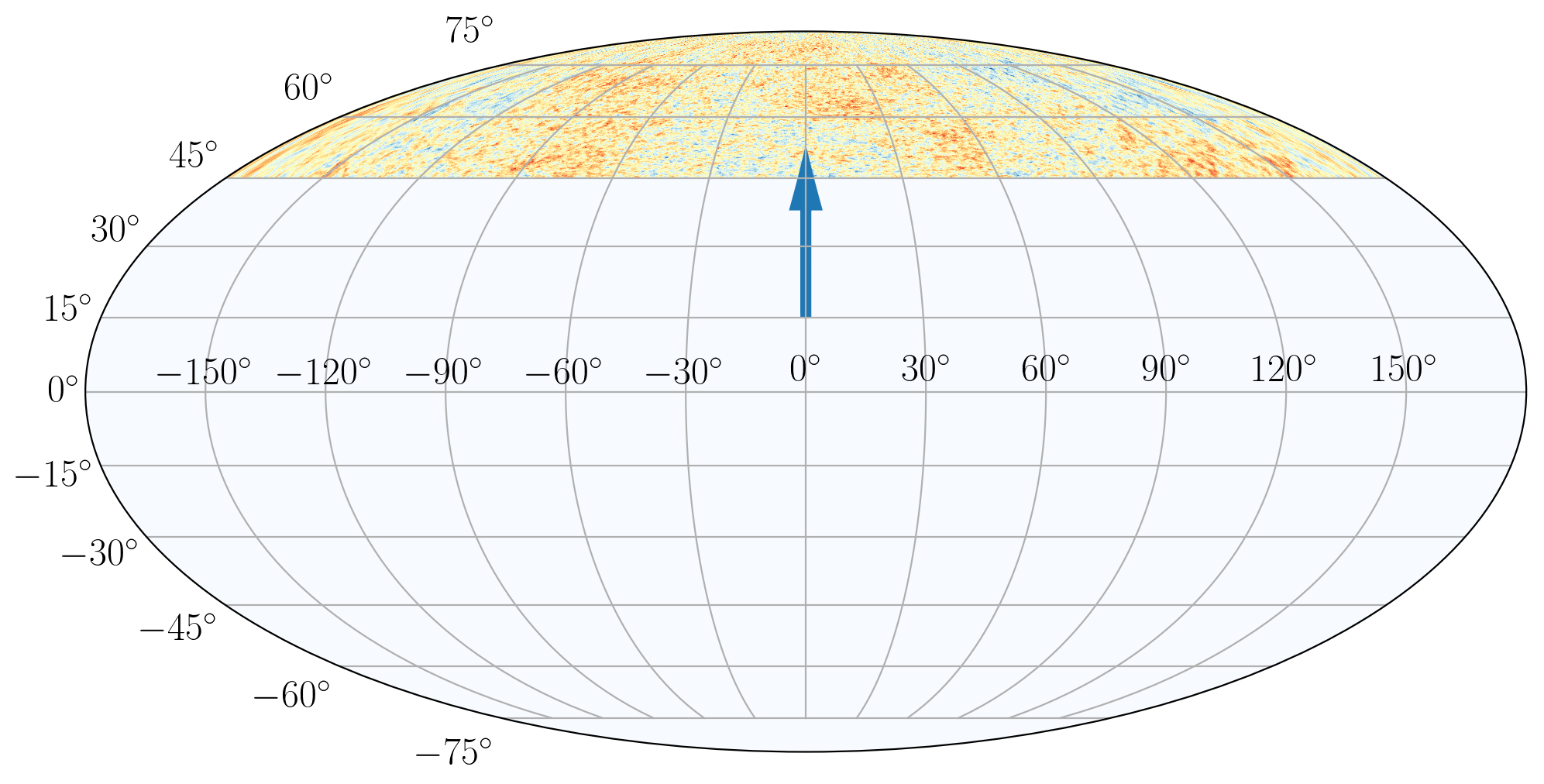}
	\caption{The theoretical cut ($b>45^\circ$, $f_\text{sky}\simeq14\%)$ applied to the CMB power spectra. The invisible part of the sky is masked and the arrow shows the direction of motion of the observer. }
	\label{fig:mask1}
\end{figure}

\subsection{Masked-Sky Power Spectra}\label{sec:IIIC}
Now we examine the effect of Doppler and aberration on a masked-sky. We will show that the motion-induced effects are more pronounced in a cut-sky compared with an all-sky, and more importantly, for specific intensity the effects increase with the frequency of observation. In any all- or partial-sky observation covering the sky symmetrically with respect to the direction of motion, the motion-induced effects on the power spectrum get suppressed, because of the cancellation between the boosted modes lying on opposite directions \cite{Dai2014}. These cancellations, however, do not occur in presence of a mask that is asymmetric with respect to the direction of motion, and hence in this case we would expect a larger effect due to  the boost. 

We compute the boosted power spectra for a representative case of a masked sky ($b>45^{\circ}$) shown in Fig.~\ref{fig:mask1} using the method outlined in Appendix A of Ref. \cite{Hivon:2001jp}. The mask has been chosen to be in a direction that maximizes the Doppler and aberration effects \cite{Jeong:2013sxy}.  Obviously, the overall power in a masked-sky is less than its all-sky analogue, so we divide the unmasked part of the sky by the average value of the mask in order to compensate the drop in the power due to the cut. This allows us to only see the drop or growth in the power due to the boost.

\begin{figure}[t]
	\centering
	Masked-sky
	\includegraphics[width=1.0\linewidth]{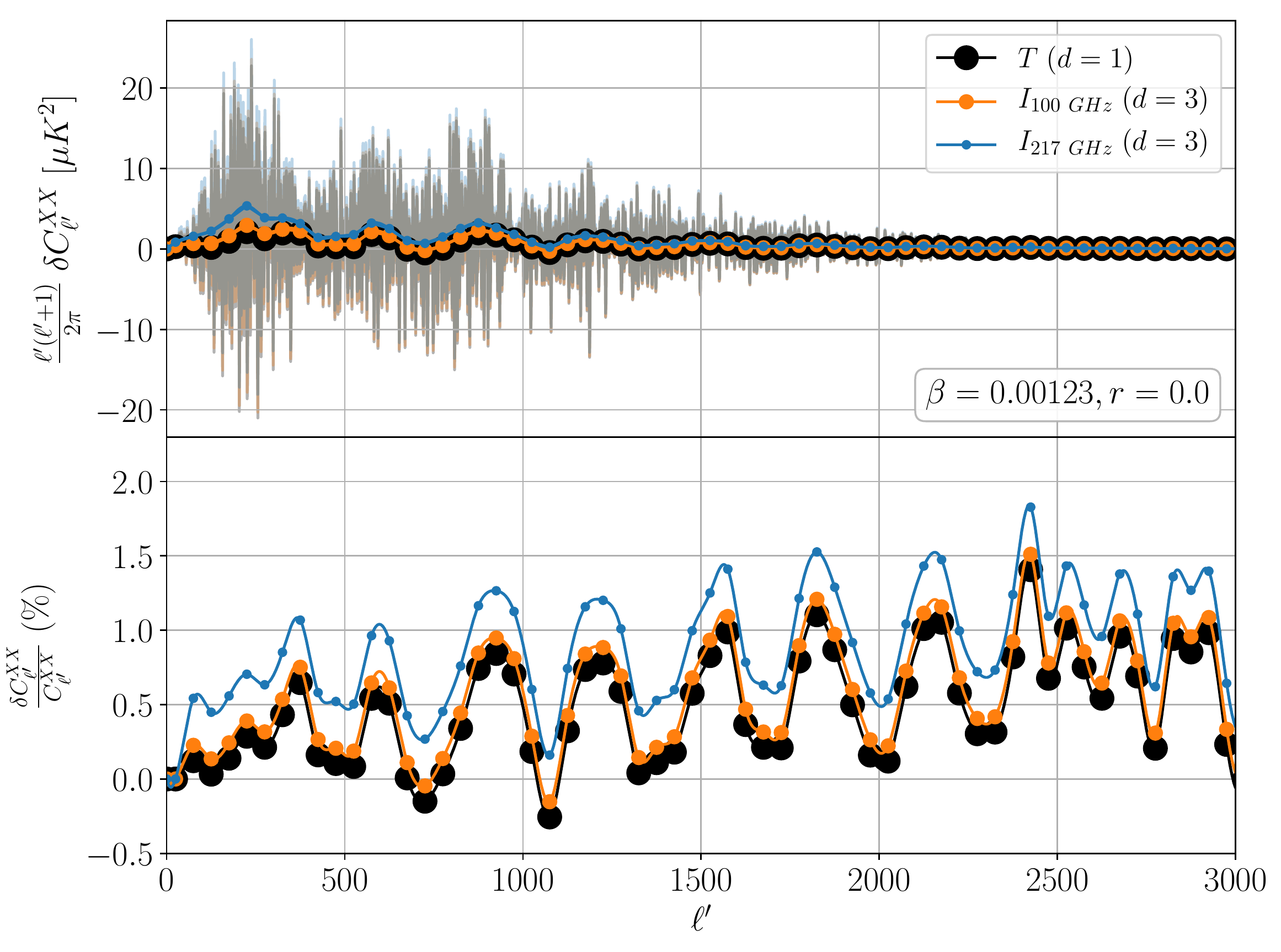}
	\caption{Equivalent of Fig. \ref{fig:TT_power_spectrum_unmasked} for a masked sky. \emph{top}: the difference between the boosted and rest frame spectra is more noticeable at lower $\ell'$ where the unpolarized CMB power spectrum is more steep (before and after the acoustic peaks and troughs). \emph{bottom}: The relative difference between the boosted and rest frame spectra in the masked-sky. The  motion-induced oscillations in temperature increase over smaller angular scales (large $\ell'$) and can reach $\sim1.5\%$. The power increase is higher at 217 GHz (100 GHz) by $\sim 0.6\% ~(0.2\%)$ over all angular scales. }
	\label{fig:TT_power_spectrum_masked}
\end{figure}

\subsubsection{Unpolarized Power Spectrum}
First, we apply the mask in Fig. \ref{fig:mask1} to a single realization of the unpolarized CMB temperature. For the all-sky map, the effect of the boost on the power spectrum was to produce small fluctuations that are more prominent where the slope of the power spectrum is at its maximum (Fig. \ref{fig:TT_power_spectrum_unmasked}). In the case of the cut-sky, these fluctuations are amplified with the mean increasing over smaller scales (Fig. \ref{fig:TT_power_spectrum_masked}). 
The increase in power in the direction of motion already exists in the all-sky map, but it is also accompanied with a decrease on power on the opposite direction of motion, so the overall effect of the boost is suppressed in this case.  However, this cancellation does not happen in a cut-sky and hence the motion-induced effects become amplified in the final observed power spectrum.

More importantly, in the case of the all-sky boosted map, there was no noticeable difference between the boosted temperature and specific intensity power spectra. However, in a masked sky the frequency-dependence of the aberration kernel becomes non-negligible. As we can see in the bottom panel of Fig. \ref{fig:TT_power_spectrum_masked}, in the masked sky, 
the difference between the power spectra of temperature and the specific intensity at 217 GHz (100 GHz) is roughly $\sim 0.6 \%$ ($\sim 0.2 \%$) over all angular scales. The difference is smaller for specific intensity at 100 GHz, as it is expected from the general behavior of the frequency-dependent aberration kernel over small frequencies. This convergence of the specific intensity aberration kernel to the temperature kernel can be seen in the single kernel elements and the BPTM for $I_\nu$ as well (Figs. \ref{fig:kernel_m_0} and  \ref{fig:BB_PS_masked}) .

\begin{figure}[t]
	\centering
	Masked-sky
	\includegraphics[width=1.0\linewidth]{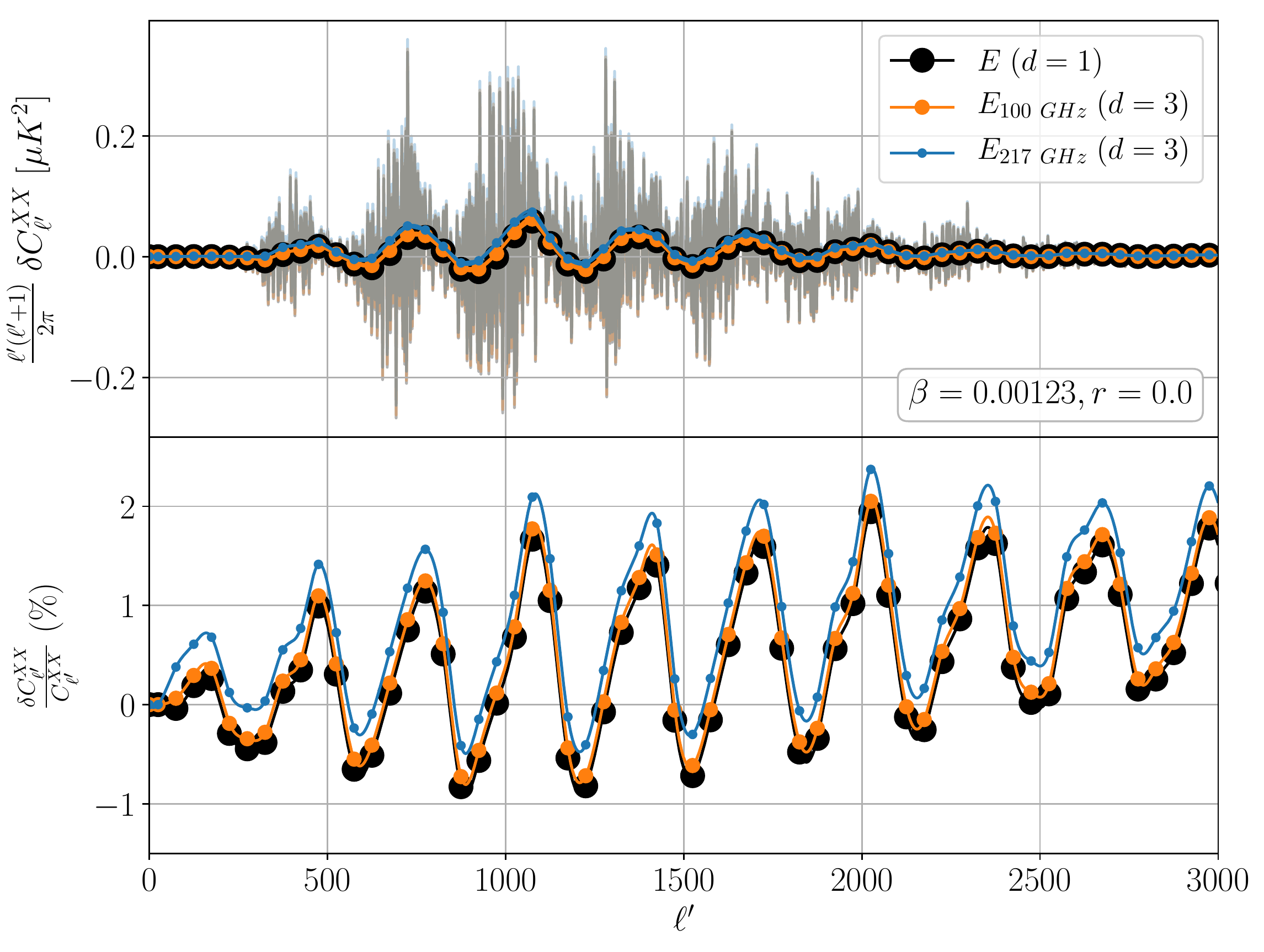}
	\caption{Equivalent of Fig. \ref{fig:TT_power_spectrum_masked} for the E mode polarization. The motion-induced oscillations are clearly pronounced before and after the acoustic peaks of the E mode power spectrum where the slope is maximum. The relative difference of the boosted and rest frame power spectra (\emph{bottom}) can be as large as $\sim2\%$. Similar to the unpolarized case, the power increase is higher at 217 GHz (100 GHz) by $\sim 0.6\% ~(0.2\%)$ over all angular scales.}
	\label{fig:EE_PS_masked}
\end{figure}

\begin{figure}[t]
	\centering
	Masked-sky
	\includegraphics[width=1.0\linewidth]{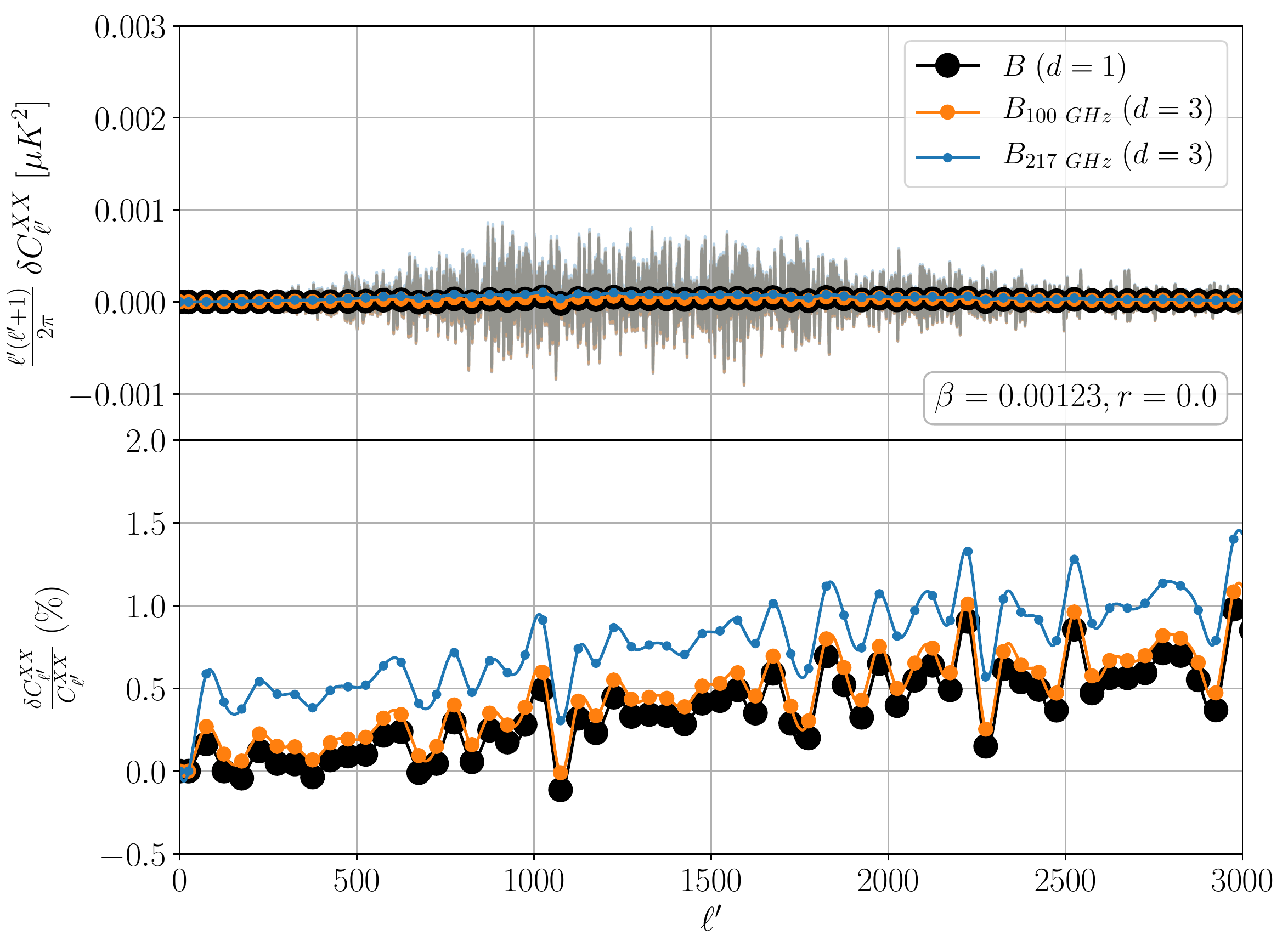}
	\caption{Equivalent of Fig. \ref{fig:EE_PS_masked} for the B mode polarization. Since the lensing power spectrum does not fluctuate over different angular scales, the motion-induced power oscillations are absent for the B mode polarization. The relative difference of the boosted and rest frame power spectra (\emph{bottom}) can be as large as $\sim1\%$. }
	\label{fig:BB_PS_masked}
\end{figure}

\subsubsection{Polarized Power Spectra}
The characteristics of the boost in the masked-sky polarized power spectra\footnote{or, more accurately, the effect of the mask on the boosted power spectra} is similar to the unpolarized case: the fluctuations due to the boost are amplified where the power spectra are steep, along with an overall increase in power over smaller angular scales. Figs. \ref{fig:EE_PS_masked} and \ref{fig:BB_PS_masked} show the effect of boost on the E and B modes of the CMB polarization power spectra. Here, the input spectra are generated with $r=0$, so the rest frame B mode polarization is only due to the lensing effect. In the top panel of Fig. \ref{fig:EE_PS_masked}, we can clearly see the prominence of the boost effect near the acoustic peaks of the E mode polarization. In contrast with the E mode, the B mode lensing power spectrum does not fluctuate over different angular scales, so the power oscillations due to the boost are absent in Fig. \ref{fig:BB_PS_masked}. 

Similar to the unpolarized power spectrum, there is an increase in power over small angular scales for both the E mode and B mode polarization. As we show in the bottom panels of Figs. \ref{fig:EE_PS_masked} and \ref{fig:BB_PS_masked}, this motion-induced power increase can respectively reach $\sim 2\%$  and  $\sim 1\%$ for the E and B modes of the polarized temperature. Here, again the power increase is larger for the specific intensity at 217 GHz (100 GHz) than the one in polarized temperature by  $\sim 0.6 \%$ ($\sim 0.2 \%$).

\begin{figure}[t]
	\centering
	Masked-sky
	\includegraphics[width=1.0\linewidth]{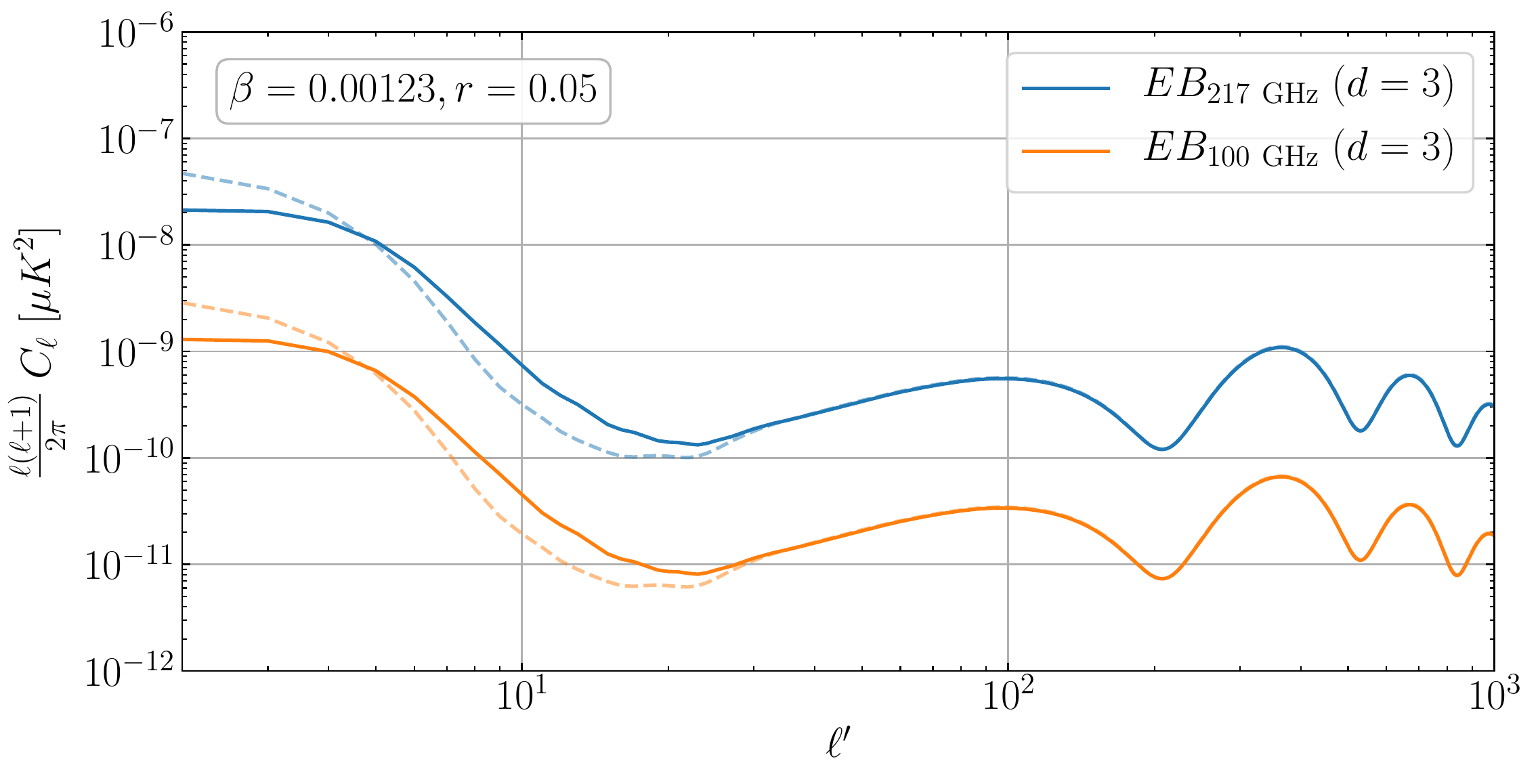}
	\caption{motion-induced E to B polarization leakage in a masked sky (\emph{solid}). The all-sky leakage spectra from Fig.  \ref{fig:polarized_power_spectrum_unmasked} are also shown for comparison (\emph{dashed}). Naturally, the mask does not affect very smaller scales ($ \ell' \gtrsim 30 $) , but there is an increase in power over large angular scales that are not cut out by the mask ($5 \lesssim \ell' \lesssim 30 $)  }
	\label{fig:EB_PS_masked}
\end{figure}

Even though the boost effects on each individual polarization mode are at the percent level, there is no significant motion-induced cross-component leakage from E to B and vice versa in the case of the masked-sky. Fig. \ref{fig:EB_PS_masked} shows the cross-leakage for E to B polarization of specific intensity for 217 and 100 GHz in a masked sky, in comparison with the all-sky case. Naturally, since our acquired cut mainly masks the large angular scales, there is no significant change in the power spectrum over small angular scales. The major modification happens at larger angular scales. For $2 \lesssim \ell' \lesssim 5 $ (corresponding to angular scales $90^\circ \lesssim \theta \lesssim 36^\circ$)  there is a drop in the power due to the fact that the mask covers a huge portion of the sky. In contrast, there is a rise in the power spectrum for $5 \lesssim \ell' \lesssim 30 $ (corresponding to angular scales $36^\circ \lesssim \theta \lesssim 6^\circ$), because of the absence of the antipodal part of the sky with respect to the direction of motion which prevents the mode cancellations to happen for these modes. For $\ell' \gtrsim 30$ the angular modes lie well within the unmasked part of the sky and hence we do not expect a major modification due to the mask for these multipoles. Nevertheless, the overall E to B leakage for polarized specific intensity seems to be small for both all-sky and masked-sky power spectra (e.g. $5\times 10^{-5} \mu$K at $\ell'=5$ for $r=0.05$). It is important to mention that the amplitude of the cross-leakage components for polarization increases with the tensor to scalar ratio $r$ and can act as a potential source of error for large scale E and B mode measurements. However, this component is easy to model and subtract from  polarization measurements of the future microwave experiments aimed at primordial gravitational waves detection through CMB polarization. 

\section{Summary}\label{sec:IV}

We introduced a frequency-dependent formalism for analyzing the motion-induced imprints of the Doppler and aberration effects in the harmonic multipoles of unpolarized and polarized radiations (\S \ref{sec:II}). Our calculations extend the formalism of DC \cite{Dai2014}---developed for boosting frequency-independent observables with arbitrary Doppler and spin weights---to include observables with arbitrary frequency spectra. This generalized formalism can be applied to CMB and other background radiations with different frequency functions. The frequency-dependent nature of the  harmonic boost equation that we introduced allows us to boost/deboost CMB with different masks or in the presence of $y$ and $\mu$ spectral distortions \cite{Sunyaev1969,Chluba:2016bvg,Balashev:2015lla} as well any type of extra galactic foregrounds.

The general effect of the boost on an observed background radiation is to generate a motion-induced leakage between the nearby observed harmonic modes ($\ell'$) [\S \ref{sec:IIIA1}]. This leakage generally is larger for closer neighbors at larger angular scales (small $\Delta \ell$ at low $\ell'$) and expands to further neighbors at smaller scales (larger $\Delta \ell$ at higher $\ell'$). For a perfect blackbody radiation the motion induced effects  increase with the frequency of observation, but they can be neglected in the Rayleigh-Jeans limit [\S \ref{sec:IIIA2}]. 

Our generalized Doppler and aberration formalism can be readily applied to CMB polarization as well [\S \ref{sec:IIA3}]. In small angular scales the spin weight of the observable becomes irrelevant and so the aberration kernel for polarized and unpolarized radiations converge. For polarized CMB observables (spin 2), the relative difference between polarized and unpolarized aberration kernel elements drop below 0.1\% for scales smaller than $\sim$ 2 degrees ($\ell \gtrsim100$). So in practice one can use the unpolarized aberration kernel elements instead of the polarized ones for these scales without much loss in precision. 

We studied the effects of a boost on the observed power spectra in all- and cut-skies for unpolarized [\S \ref{sec:IIIB3}] and polarized [\S \ref{sec:IIIB4}] temperature, integrated intensity and specific intensity at 100 and 217 GHz. In the all-sky case, the boost produces small flat oscillations ($\sim ~0.1 \%$) over all angular scales and we showed that the frequency of observation is practically irrelevant for a pure blackbody CMB spectrum for both polarized and unpolarized maps. When using polarized thermodynamic temperature, there is no leakage from E to B mode and vice versa. However, we show that with polarized specific intensity, there is a cross-component leakage which increases with the observational frequency. Even though the E to B leakage is larger at 217 GHz compared with the one at 100 GHz, they are both largely subdominant to the primordial E and B modes (for $r=0.05$) and their individual boost corrections. This leakage is of the order of a few $\times~10^{-5} \mu$K for the  specific intensity (normalized to temperature units) over large angular scales $(\ell'>5)$ and declines over smaller scales. Applying a mask on the power spectrum does not greatly affect the cross-component leakage and the overall effect remains sub-dominant with respect to the CMB intrinsic E and B modes. Nevertheless, using the formalism introduced in this paper, the motion-induced effects on individual E and B  modes and their respective cross-component leakages can be easily modeled and subtracted from observations, along with other systematic errors in experiments aiming at measuring the gravitational waves through CMB. 

The effects of the Doppler and aberration in the direction of motion (e.g. increase in frequency of the photons or decrease in the solid angle) are generally accompanied by the opposite effects on the antipodal direction on the sky. The overall impression of the boost on the all-sky power spectrum for an individual angular mode is therefore canceled due to averaging over modes lying on opposite sides of the sky (mathematically speaking, due to averaging over different $m$ modes of a perticular $\ell'$ mode). This cancellation, however, does not happen for a cut-sky, where the motion-induced effects are especially larger when the mask is asymmetric with respect to the direction of motion [\S \ref{sec:IIIC}]. For the specific cut ($b>45^\circ$, $f_\text{sky}\simeq14\%)$ that we applied to a single sky realization, the boost can increase both unpolarized and polarized (E-mode) thermodynamic temperature by 1-2\% at angular scales smaller than 10 arcmins ($\ell'\gtrsim1000$). The motion-induced effects are typically larger near the angular scales where the slope of the rest frame power spectra are most steep \cite{Jeong:2013sxy}. Since the B mode polarization power spectrum lacks this feature (at least on small scales where lensing is the dominant source), the motion-induced effects for this mode are not as prominent as the ones in temperature and E mode polarization, reaching at most $1\%$ at 4 arcmins ($\ell'=2500$). 

More importantly, the frequency of observation for specific intensity becomes relevant in a cut-sky. We showed that for both unpolarized and polarized power spectra the motion-induced effects for specific intensity at 217 GHz (100 GHz) are larger than their thermodynamic temperature counterparts by $\sim 0.6\%$ 
($\sim 0.2\%$) over all angular scales. Therefore, in principle neglecting the frequency dependence of the boost can generate a bias in cosmological parameter estimation from the power spectra. 
 
It is important to mention that all the reported numbers are specific to the mask that we applied to the sky realizations and power spectra, and are only valid for an observer moving in the north galactic pole with the velocity $\beta =0.00123$. These examples were only chosen to present the importance of the frequency dependence of the generalized Doppler and aberration kernel that we developed in this paper. A more detailed analysis of the results on the CMB power spectra with realistic masks, window functions and proper local direction of motion for the observer is therefore needed to accurately gauge the amplitude of the motion-induced effects for different CMB experiments. 

\section{Acknowledgments} 
The authors thank Loris Colombo and Nareg Mirzatuny for helpful discussions. We are sincerely grateful to the anonymous referee for their exceedingly detailed and helpful comments on our manuscript. EP is supported by the WiSE major faculty funding. SY is supported by the USC graduate final year fellowship. Some of the results in this paper have been derived using the HEALPix/healpy \cite{Healpix} package \footnote{ \href{http://healpix.sourceforge.net}{http://healpix.sourceforge.net}}.

\bibliographystyle{apsrev4-1}	
\bibliography{kSZPol}
\end{document}